\documentclass[11pt]{article}
\usepackage{epsf,rotating,latexsym,astron}
\newcommand{\mc}[3]{\multicolumn{#1}{#2}{#3}}
\begin{document}
\bibliographystyle{astron}

\begin{center}
{\Large \bf One year of Galileo dust data from the Jovian system: 1996}
\end{center}

\bigskip

{\bf
        H.~Kr\"uger$^a$\footnote{{\em Correspondence to:} 
Harald.Krueger@mpi-hd.mpg.de}, 
        E.~Gr\"un$^a$,
        A.~Graps$^a$,
        D.~Bindschadler$^b$,
        S.~Dermott$^c$,
        H.~Fech\-tig$^a$,
        B.~A.~Gustaf\-son$^c$,
        D.~P.~Hamilton$^d$, 
        M.~S.~Hanner$^b$,
        M.~Hor\'anyi$^e$,
        J.~Kissel$^f$,
        B.~A.~Lind\-blad$^g$,
        D.~Linkert$^a$,
        G.~Linkert$^a$,
        I.~Mann$^{h,i}$,
        J.~A.~M.~McDonnell$^j$,
        G.~E.~Mor\-fill$^f$, 
        C.~Polanskey$^b$,
        G.~Schwehm$^{i}$,
        R.~Srama$^a$, 
        and H.~A.~Zook$^{k}$\footnote{Passed away on 14 March 2001}
}

\bigskip

\small
\begin{tabular}{ll}
a)& Max-Planck-Institut f\"ur Kernphysik, 69029 Heidelberg, Germany\\
b)& Jet Propulsion Laboratory, Pasadena, California 91109, USA\\
c)& University of Florida, Gainesville, FL\,32611, USA \\
d)& University of Maryland, College Park, MD\,20742-2421, USA\\
e)& Laboratory for Atmospheric and Space Physics, Univ.
                 of Colorado, Boulder, \\ 
  & CO\,80309, USA\\
f)& Max-Planck-Institut f\"ur Extraterrestrische Physik, 85748 Garching, 
                                                                   Germany\\ 
g)& Lund Observatory, 221 Lund, Sweden\\
h)& Institut f\"ur Planetologie, Universit\"at M\"unster, 48149 M\"unster, Germany\\
i)& ESTEC, 2200 AG Noordwijk, The Netherlands\\
j)& Planetary and Space Science Research Institute, The Open University, \\
  & Milton Keynes, MK7 6AA, UK\\
%10)& European Space Research and Technology Center, 2200 AG Noordwijk, 
%                                                      The Netherlands\\
k)& NASA Johnson Space Center, Houston, Texas 77058, USA\\
\end{tabular}

\normalsize

\bigskip

\begin{abstract}

The dust detector system onboard Galileo records dust impacts in circumjovian 
space since the spacecraft has been injected into a bound orbit about Jupiter 
in December 1995.
This is the sixth in a series of papers dedicated to presenting Galileo and 
Ulysses dust data. We present data from the Galileo dust instrument for the 
period January to December 1996 when the spacecraft completed four orbits about Jupiter
(G1, G2, C3 and E4). Data were obtained as high resolution realtime science data 
or recorded data during a time period of 100 days, or via memory read-outs 
during the remaining times.
Because the data transmission rate of the spacecraft is very low, the 
complete data set  
(i.~e. all parameters measured by the instrument during impact of a dust 
particle) for only  2\% (5353) of all particles detected could be transmitted 
to Earth; the other particles were only counted. Together with the data for 
2883 particles detected during Galileo's interplanetary cruise and published 
earlier, complete data of 8236 particles detected by the Galileo
dust instrument from 1989 to 1996 are now available. 
The majority of particles detected are tiny grains (about 10 nm in radius) 
originating from Jupiter's innermost Galilean moon Io. These grains have 
been detected throughout the Jovian system and the highest impact rates
exceeded $\rm 100\, min^{-1}$. A small number of grains has been 
detected in the close vicinity of the Galilean moons Europa, Ganymede
and Callisto which belong to impact-generated dust clouds formed by 
(mostly submicrometer sized) ejecta from the surfaces of the moons 
(Kr\"uger et al., Nature, 399, 558, 1999). Impacts of submicrometer 
to micrometer sized grains have been detected thoughout the Jovian 
system and especially in the region between the Galilean moons.

\end{abstract}

\section{Introduction}

In December 1995, Galileo became the first artifical satellite to 
orbit Jupiter. 
The main goals of the Galileo mission 
are the exploration of the giant planet itself, its four 
Galilean satellites and its huge magnetosphere. Galileo
carries a highly sensitive impact ionization dust detector on 
board, which is a twin of the dust detector on the Ulysses 
spacecraft \cite{gruen1992a,gruen1992b,gruen1995a}. 
Dust data from 
both spacecraft have been used many times for the analysis of 
e.\,g. the interplanetary dust complex, 
dust related to asteroids and comets, 
interstellar dust grains sweeping through the solar system, 
and streams of dust particles originating from the 
Jupiter system. Here we recall only publications which
are related to dust in the Jupiter system.
See Kr\"uger et al. \cite*{krueger1999a,krueger1999b} for a summary of references to 
other works.

Streams of dust particles originating from the Jovian system have 
been discovered with the Ulysses detector \cite{gruen1993} and
later been confirmed by Galileo \cite{baguhl1995a,gruen1996b}. 
The first dust detection with Galileo in the Jupiter system itself was 
reported from Galileo's initial approach to Jupiter and Io fly-by in 
December 1995 \cite{gruen1996c}. 
At least three different types of dust particles have been 
identified in the Jupiter system \cite{gruen1997b,gruen1998}: 
1. Streams of dust particles with high and variable impact rates 
throughout Jupiter's magnetosphere. The particles are 
about 10\,nm in diameter \cite{zook1996,liou1997}
and they originate from the 
innermost Galilean satellite Io \cite{krueger1998,graps2000}. Because
of their small sizes the particles strongly interact with Jupiter's 
magnetosphere \cite{horanyi1997,heck1998}. These 
streams of particles have been used to analyse the field of 
view of the dust detector \cite{krueger1999b} at a level of 
detail which could not be achieved during ground calibration.
2. Sub-micrometer grains which form dust clouds surrounding the Galilean
moons \cite{krueger1999d,krueger2000}. These grains are ejected from the 
satellites' surfaces by hypervelocity impacts of interplanetary dust particles.
3. Bigger micrometer-sized grains  forming a tenuous dust ring 
between the Galilean satellites. This third group is composed of two 
sub-populations, one orbiting Jupiter on prograde orbits and a second one on 
retrograde orbits \cite{colwell1998,thiessenhusen2000}. Most of the 
prograde population is maintained by grains escaping
from the clouds surrounding the Galilean satellites \cite{krivov2001}.

This is the sixth paper in a series dedicated to presenting both 
raw and reduced data from the Galileo and Ulysses dust instruments. 
The reduction process of Galileo and Ulysses dust data has been 
described by Gr\"un et al. \cite*{gruen1995a} (hereafter Paper~I).
In Papers~II and IV \cite{gruen1995b,krueger1999a} 
we present the Galileo data set spanning the 
six year time period October 1989 to December 1995. Papers~III 
and V \cite{gruen1995c,krueger1999b} discuss the five 
years of Ulysses data from October 1990 to December 1995.
The present paper extends the 
Galileo data set from January to December 1996, which is part of 
Galileo's prime Jupiter mission. A companion paper \cite{krueger2001b}
(Paper~VII) details Ulysses' measurements from 1996 to 1999.

The main data products are a table of the number of all impacts determined 
from the accumulators of the dust instrument and a table of both raw and
reduced data of all ``big'' impacts received on the ground (the term 
``big'' applies to impacts in ion amplitude ranges AR2 to AR6; see 
Section~3 in this paper and Paper~I for a definition of the amplitude 
ranges). The information presented in this paper is similar to data 
which we are submitting to the various data archiving centers (Planetary 
Data System, NSSDC). The only difference is that the paper version 
does not contain the full data set of the large number of ``small'' 
particles (amplitude range AR1), and the numbers of impacts deduced from 
the accumulators are typically averaged over one to several days. 
Electronic access to the 
complete data set including the numbers of impacts deduced from the 
accumulators in full time resolution is also possible via 
the world wide web: http://www.mpi-hd.mpg.de/dustgroup/.

This paper is organised similarly to our previous papers. Section~\ref{mission} 
gives a brief overview of the Galileo mission until the end of 1996,
the dust instrument and lists important mission events in 1996.
A description of the new Galileo dust data set for 1996 
together with a discussion of the detected noise and dust impact rates is given 
in Sect.~\ref{events}. Section~\ref{analysis} analyses and discusses 
various characteristics of the new data set. 
Finally, in Sect.~\ref{discussion} we discuss
results of the Jovian dust complex achieved with the new data set.

\section{Mission and instrument operations} \label{mission}

Galileo was launched on 18 October 1989. Two fly-bys at 
Earth and one at Venus between 1990 and 1992 gave the spacecraft 
enough energy to leave the inner solar system. During its 
interplanetary voyage Galileo had close encounters with the 
asteroids Gaspra and Ida. On 7 December 1995 the spacecraft 
arrived at Jupiter and was injected into a highly elliptical orbit 
about the planet. Galileo's trajectory during its orbital tour about 
Jupiter from December 1995 to early January 1997 is shown in 
Fig.~\ref{trajectory}. Galileo
now performs regular close fly-bys of Jupiter's Galilean satellites.
Four such encounters occurred in 1996 (two at Ganymede, one at Callisto
and one at Europa, see Tab.~1).
%\ref{event_table}). 
Galileo orbits 
are labelled with the first letter of the Galilean 
satellite which was the encounter target during that orbit,
followed by the orbit number. For example, 
``E4'' refers to Galileo's fourth orbit about Jupiter which had a 
close fly-by at Europa. Satellite fly-bys occurred within two days of 
Jupiter closest approach (pericenter passage). A detailed description of 
the Galileo mission and the spacecraft have been given by Johnson et al. 
\cite*{johnson1992} and D'Amario et al. \cite*{d'amario1992}. 

\begin{center} \fbox{\bf Insert Fig.~\ref{trajectory}} \end{center}

\begin{figure}
\epsfxsize=15cm
\epsfbox{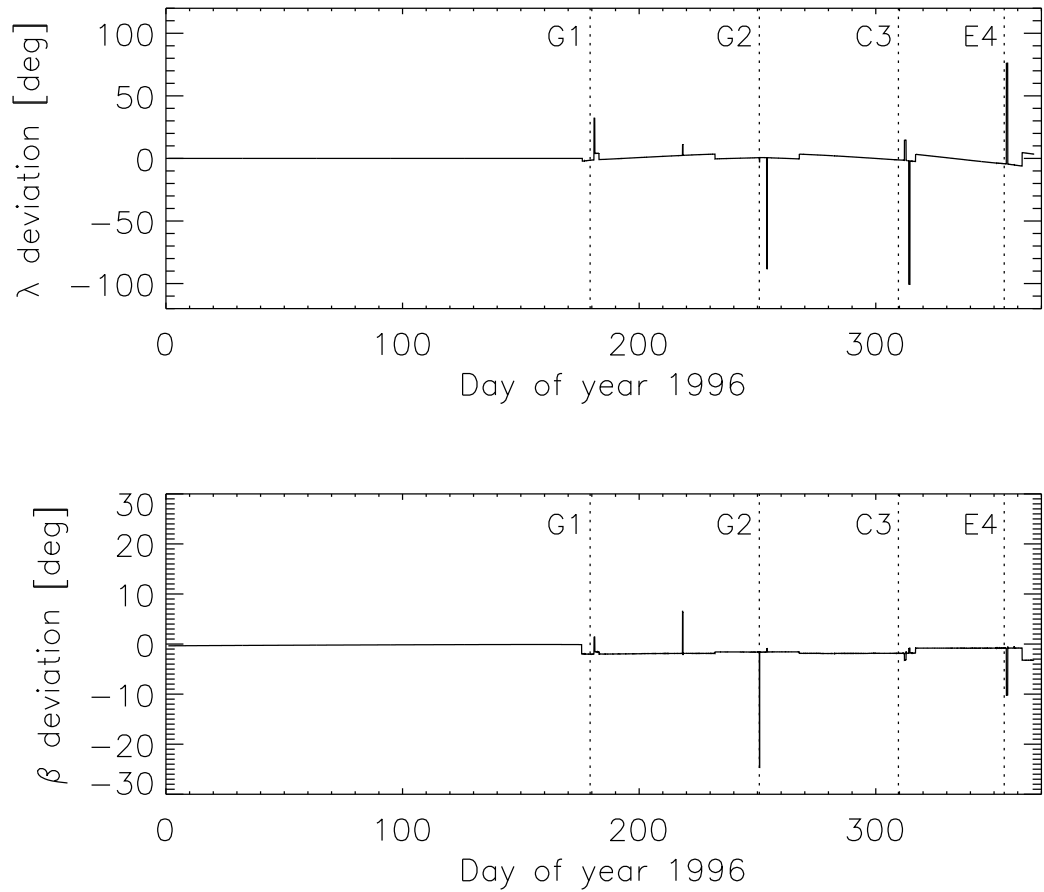}
%\epsfbox{../bilder/pointing.ps}
        \caption{\label{pointing}
Spacecraft attitude: deviation of the antenna pointing direction 
(i.~e. negative spin axis) from the Earth direction. The angles are 
given in ecliptic longitude (top) and latitude (bottom, equinox 1950.0)
The four targeted encounters of Galileo with the Galilean satellites 
are indicated by dotted lines.
Sharp spikes are associated with imaging observations with 
Galileo's cameras or orbit trim maneuvers with the spacecraft thrusters.
}
\end{figure}

Galileo is a dual spinning spacecraft with an antenna that points 
antiparallel to the positive spin axis. During most of the initial 3 years 
of the mission the antenna pointed towards the Sun (see 
Paper~II). Since 1993 the antenna has usually been pointing towards Earth.
Deviations from the Earth pointing direction 
in 1996, which is the time period considered 
in this paper, are shown in Fig.~\ref{pointing}.	
Sharp spikes in the pointing deviation occurred when the spacecraft was turned
away from the nominal Earth direction for dedicated imaging observations with 
Galileo's cameras or for orbit trim maneuvers with the spacecraft thrusters.
These spikes lasted typically several hours (see Tab.~1).

\begin{center} \fbox{\bf Insert Fig.~\ref{pointing}} \end{center}

The Dust Detector System (DDS) is mounted on the spinning section of 
Galileo and the sensor axis is offset by an angle of $60^{\circ}$
from the positive spin axis (an angle of $55^{\circ}$
has been erroneously stated earlier). A schematic view 
of the Galileo spacecraft and the geometry of dust detection
are shown in the inset of Fig.~\ref{trajectory} (see also Paper~IV and 
Gr\"un et al. \cite*{gruen1998}).

The rotation angle measures the viewing 
direction of the dust sensor at the time of a dust impact. 
During one spin revolution of the spacecraft the rotation angle scans 
through a complete circle of $360^{\circ}$. At rotation angles of 
$90^{\circ}$ and $270^{\circ}$ the sensor axis lies nearly 
in the ecliptic plane, and at $0^{\circ}$ it is close to the ecliptic 
north direction. 
DDS rotation angles are taken positive around the negative spin axis of 
the spacecraft. This is done to easily compare Galileo spin 
angle data with those taken by Ulysses, which, unlike
Galileo, has its positive spin axis pointed towards Earth.
DDS has a $140^{\circ}$ wide field of view, although a smaller 
field of view applies to a subset of dust impacts -- the so-called class~3 
impacts \cite{krueger1999c}. During one spin revolution
of the spacecraft the sensor axis scans a cone with $120^{\circ}$
opening angle towards the anti-Earth direction. Dust particles which arrive
from within $10^{\circ}$ of the positive spin axis (anti-Earth direction) 
can be detected at all rotation angles, whereas those that arrive at angles 
from $10^{\circ}$ to $130^{\circ}$ from the positive spin axis can 
be detected over only a limited range of rotation angles.

In June 1990 DDS was reprogrammed for 
the first time after launch and since then the DDS memory can store 46 
instrument data frames (with each frame comprising the
complete data set of an impact or noise event, consisting of 128 bits,
plus ancillary and engineering data; see~Papers~I and II). DDS time-tags
each impact event with an 8 bit word allowing for the identification of 
256 unique times. Since 1990 the step size of this time 
word is 4.3~h. The total accumulation time after 
which the time word is reset and the time labels of older impact events 
become ambiguous is $\rm 256 \times 4.3\,h = 46\,$days.  

During most of the interplanetary cruise (i.\,e. before 7 December 
1995) we received DDS data as instrument memory-readouts (MROs).
MROs return event data which have accumulated in the 
instrument memory over time. The contents of all 46 instrument data 
frames of DDS is transmitted to Earth during an MRO. 
%The complete data set of only 40 different events (dust impacts
%or noise events) is contained in an MRO because the information of the 
%latest event in each 
%of the six amplitude ranges (Paper~I) is stored twice (the so-called
%A range). 
If too many events occur between two MROs, the data sets of the oldest 
events become overwritten in the instrument memory and are lost. 

Because the high-gain antenna of Galileo did not open completely, the 
on-board computer of the spacecraft had to be reprogrammed to establish a completely
new telecommunications link. New flight-software was installed in the 
spacecraft computer in April and May 1996 
\cite{statman1997} which provided a new mode for high-rate dust data 
transmission to the Earth, the so-called realtime science mode (RTS). 
In RTS mode, DDS data were read-out either every 7.1 or every 21.2 minutes, 
depending on the spacecraft data transmission rate, and were usually directly 
transmitted to Earth with a rate of 3.4 or 1.1 bits per second, respectively.
%Only during short time intervals RTS data 
%were stored on Galileo's tape recorder and transmitted to Earth later.  
For short periods (i.e. $\rm \sim \pm 1/2\,h$) around satellite 
closest approaches, DDS data were collected with a higher rate of about 24 
bits per second, recorded on the tape recorder 
(record mode) and transmitted to Earth several days to a few weeks later. 
Sometimes RTS data for short time intervals were also stored on the tape recorder
and transmitted later but this did not change the labelling -- they are 
called RTS.  

In RTS and record mode only seven instrument data frames are read-out at a 
time and transmitted to Earth rather than the complete instrument memory.  Six 
of the frames contain the information of the six most recent events in 
each amplitude range (see~Paper~I and Section~\ref{events} for a definition of
the amplitude ranges). The seventh frame belongs to an older event 
read-out from the instrument memory (FN=7) and is transmitted in
addition to the six new events. The position in the instrument memory
from which this seventh event is read changes for each read-out so that after
40 read-outs the complete instrument memory gets transmitted (note that the 
contents of the memory may change significantly during the time period of 40 
read-outs if high event rates occur). 
Although fewer data frames 
can be read-out this way at a time, the number of new events that can be 
transmitted to Earth in a given time period is much larger than with MROs 
because the read-outs occur much more frequently.

In 1996, RTS and record data were obtained during a period of 100 days 
(Fig.~\ref{trajectory}). 
During the remaining times when DDS was  operated in neither RTS nor
record mode, MROs occurred at several day intervals. Except before 
29 March 1996 the MROs were frequent enough so that no ambiguities 
occurred in the time-tagging (i.e. MROs occurred at intervals smaller than
46 days). 

\begin{center} \fbox{\bf Insert Tab.~1} \end{center}

Table~1
%\ref{event_table} 
lists significant mission and dust instrument 
events for 1996. A comprehensive list of earlier 
events can be found in Papers~II and IV. After Galileo's Io and 
Jupiter fly-bys on 7 December 1995, the channeltron high voltage of 
DDS was switched off completely. More than three months later, on 
27 March 1996, the instrument was reactivated and brought into the 
same nominal mode with which it was operated during most of Galileo's
interplanetary cruise to Jupiter: the channeltron voltage was set to 
1020~V (HV~=~2), 
the event definition status was set such that the channeltron or 
the ion-collector channel can independently initiate a measurement cycle 
(EVD~=~C,I) and the detection thresholds for the charges on the 
ion-collector, channeltron, 
electron-channel and entrance grid were set (SSEN~=~0,~0,~1,~1). This 
was also the nominal configuration for most of 
the orbital tour in the Jovian system.
Detailed descriptions of the symbols are given in Paper~I. 

\begin{figure}
\vspace{-2cm}
\epsfxsize=15cm
\epsfbox{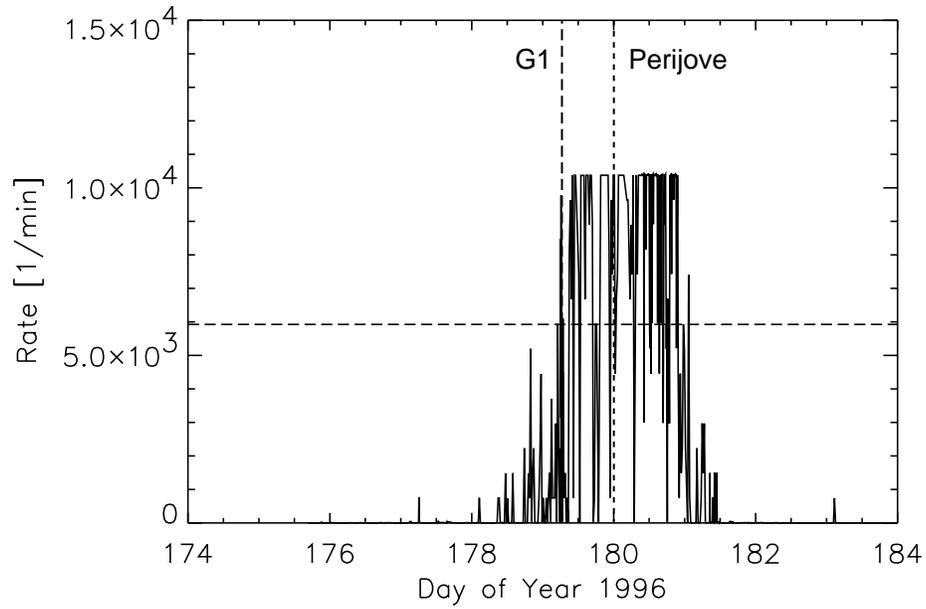}
%\epsfbox{../bilder/g1noise.ps}
\vspace{-4.5cm}
        \caption{\label{g1noise}
Channeltron noise rate during the G1 orbit in the inner Jovian system. The 
noise rate of the smallest
class 0 events (AR1, positive signal charge $\rm Q_{\,I} < 10^{-13}\,C$) plus
the noise rate detected by the channeltron noise counter are shown. The dashed
horizontal line indicates noise rates above which considerable dead-time
occurs. The vertical dashed lines indicate the G1 Ganymede fly-by and perijove 
passage (perijove distance from Jupiter was $\rm 11\,R_J$).
}
\end{figure}

When Galileo performed its first passage through the inner 
Jovian system after insertion into an orbit about Jupiter (G1 orbit, 
June 1996) DDS was operated in its nominal 
mode as described above. This was also the mode DDS was operated in during
Galileo's 6 year interplanetary cruise. In this mode high channeltron noise 
rates were recorded within about 20$\rm R_J$ distance from Jupiter 
(Jupiter radius, $\rm R_J = 71,492\,km$) 
which reached values of up to 10,000 events per minute (Fig.~\ref{g1noise}). 
Because the time the onboard computer of DDS needs to process data from
a single event (impact or noise) is about 10 msec \cite{gruen1995a},
significant dead time is produced when the event rate exceeds 6,000 
per minute.  Hence, only very few dust impacts could be recorded from 
day 179.5 to 181.0 in 1996.
%From day 181 to 183, when the noise
%rate was low again, no small dust impacts were detected.
During all later orbits of Galileo's prime Jupiter mission the event 
definition status and the detection threshold for the channeltron charge 
were changed within 18$\rm R_J$ (see Tab.~1).
%\ref{event_table}). 
This reduced the noise sensitivity in the inner Jovian system and effectively 
prevented dead-time problems in the G2 and all later orbits.

\begin{center} \fbox{\bf Insert Fig.~\ref{g1noise}} \end{center}

During the Jupiter orbital tour of Galileo, orbit trim maneuvers 
(OTMs) have been 
executed around perijove or apojove passages to target
the spacecraft to close encounters with the Galilean 
satellites. A few of these maneuvers required changes in the 
spacecraft attitude off the nominal Earth pointing direction
(see Fig.~\ref{pointing}). 
In one case, on 9 September 1996, the spacecraft was turned by 
$\rm 88^{\circ}$. In addition, dedicated spacecraft turns occurred 
typically in the inner Jovian system within a few days around  
perijove to allow for imaging observations with Galileo's cameras or
to maintain the nominal Earth pointing direction. A large turn by
$\rm 99^{\circ}$ off the Earth direction occurred  on 8 November 1996. 
Both attitude changes on 9 September and 8 November were large enough that 
DDS could record impacts of dust stream particles at times
when these particles would have been undetectable with the 
nominal spacecraft orientation \cite{gruen1998}. 

A spacecraft anomaly  occurred on 24 August 1996 and no dust data were
obtained until 31 August when the spacecraft was recovered. Although 
the dust instrument continued to collect data they could not be transmitted 
to Earth and most of them were lost. The data set for only 5 grains 
detected in this time period was transmitted as data read from the instrument
memory (FN=7).

DDS classifies and counts all registered events with 24 dedicated accumulators
(see Section~\ref{events} and Paper~I). During Galileo's first orbits about Jupiter 
(G1, G2 and C3) unanticipated high rates occurred in two of the highest quality 
categories and unrecognized accumulator overflows may have occurred. 
To detect such overflows, DDS was reprogrammed on 4 December 1996 
by adding two overflow counters. 
%Since then all detected particles have been 
%recognized with the accumulators and no unrecognized overflows 
%occurred. 
The details of this reprogramming and the significance of possible
accumulator overflows are described in Section~\ref{events}.

\section{Impact events} \label{events}     

DDS classifies all events -- real impacts of dust particles and noise 
events -- into one of 24 different categories (6 amplitude ranges for
the charge measured on the ion collector grid 
and 4 event classes) and counts them in 24 corresponding 8 bit accumulators 
(Paper~I).  In interplanetary space most of the 24 categories described above were 
relatively free from noise and only sensitive to real dust impacts, 
except for extreme situations like the crossings of the radiation 
belts of Earth, Venus (Paper~II) and Jupiter (Paper~IV). During most 
of Galileo's initial three years of interplanetary cruise only the lowest 
amplitude and class categories -- AC01 (event class 0, amplitude 
range 1, AR1), AC11, and AC02 -- were contaminated by noise events 
(Paper~II). In July 1994 the onboard classification scheme 
of DDS was changed to identify smaller dust impacts in the data. 
With the modified scheme noise events 
were usually restricted to class 0 but may have occurred in all amplitude ranges. 
All events in higher quality 
classes detected in the low radiation environment of interplanetary 
space were true dust impacts 
(class~0 may still contain unrecognized dust impacts). 

In the extreme radiation environment of the Jovian system, a 
different noise behaviour of the instrument was recognized: 
especially when Galileo was within about 20 $\rm R_J$ 
from Jupiter the higher event classes were contaminated by noise 
(see also Paper~IV). This noise, which affects 
class~1 and class~2, is unrelated to the channeltron noise
shown in Fig.~\ref{g1noise}. In an analysis of the whole 
dust data set from Galileo's prime Jupiter mission, noise events
could be eliminated from class~2 \cite{krueger1999c}. 
Class~1 events, however,  show signatures of
being nearly all noise events in the Jovian environment. We therefore consider
the class~3 and the denoised class~2 
impacts  as the complete set of dust data from Galileo's Jupiter tour.
Apart from a missing third charge signal -- class~3 has three charge 
signals and class~2 only two -- there is no physical difference between
dust impacts categorized into class~2 or class~3. 

In this paper the terms ''small`` and ''big`` have the
same meaning as in Paper~IV (which is different from the terminology 
of Paper~II). We call all particles in class~2 and class~3 in the 
amplitude ranges 2 and higher (AR2 to AR6) ''big``. Particles in
the lowest amplitude range (AR1) are called ''small``. This distinction
separates the small Jovian dust stream particles from bigger grains 
which are mostly detected between and near the Galilean satellites.

In RTS and record mode the time between two readouts of the 
instrument memory determines the number of events in a given time period 
for which their complete information can be transmitted.
Thus, the complete information on each impact is
transmitted to Earth when the impact rate is below one impact per
either 7.1 or 21.2 minutes in RTS mode or one impact per minute in
record mode.  If the impact rate exceeds these 
values, the detailed information of older events is lost because 
the full data set of only the latest event is stored in the DDS
memory. 

Furthermore, in these two modes the time between two read-outs
also limits the accuracy with which the impact time can be determined. 
It is 7.1 or 21.2 minutes in RTS mode and about one minute in record mode, 
respectively. 
%During MROs 
%the complete instrument memory is
%read-out and  the information of all 40 impacts is transmitted to Earth.
%If too many impacts occur between two MROs the detailed information of the
%older particles is lost.
During times when only MROs occurred, the  
accuracy is limited by the increment of the DDS internal clock, i.e. 4.3~hours. 

Because of the large differences in the timing accuracy in the various 
read-out modes, we have defined a new parameter, time error value -- TEV,
that determines the accuracy of the impact time of a dust particle in 
minutes. TEV has been rounded to the next higher integer. In RTS and
record mode, TEV is usually simply the time between two read-outs, i.e.
TEV = 8 or 22 min in RTS or TEV = 2 min in record mode, respectively. During
gaps in the data transmission, i.e. when data packets were lost, multiples 
of these values occur. Usually, the given impact time of a dust particle
is identical with the readout time of the instrument, which means that the
impact has occurred some time in the period impact time minus TEV. This
is the case for almost all impacts in AR2 to AR6 and more than 70\% of 
the impacts in AR1. For example the impact of particle 3497 in Tab.~5 
(TEV = 8 min) occurred between 96-180 10:06 and 96-180 10:14.

Data frames belonging to older events not transmitted immediately after impact
but transmitted later (as the seventh event from each read-out, FN=7)
have impact times interpolated to lie between the times of the two adjacent 
read-outs. Their TEV value is equal to the time interval between
these adjacent read-outs. For MROs the impact time has been interpolated to the
middle of the time interval defined by the DDS internal clock. The 
corresponding TEV is 259 min $\rm \sim 4.3\,h$. For example, impact 3780 
in Tab.~5 (TEV = 259 min) occurred at 96-241 23:55 $\rm \pm 259/2$ min.

DDS records and counts the number of all dust particle
impacts and noise events with 8 bit accumulators. The time between two
readouts of the instrument memory determines the maximum rate
which can unambiguously be derived from the accumulators. At rates 
below $\rm 256\,/\,
7.1\,\sim 36\,min^{-1}$ or $\rm 256\,/\, 21.2\,\sim 12\,min^{-1}$
in RTS mode and below $\rm \sim 200\,min^{-1}$ 
in record mode the accumulator values transmitted to Earth represent the 
true event rates. During times of higher event rates an unknown number 
of accumulator overflows occurred which led to ambiguities in the number 
of events derived from the accumulators. Thus, the derived rates may 
be underestimated. 

To cope with unanticipated high rates in the inner Jovian 
system (see Fig.~\ref{rate} and Gr\"un et al. \cite*{gruen1998}) 
DDS  was reprogrammed on 4 December 1996. Two of the 
24 accumulators (AC05 and AC06) were modified to count the number of 
accumulator overflows of the two highest quality classes in the lowest 
amplitude range (AC21 and AC31). Since then the
maximum recordable  rate in RTS mode is $\rm 256^2\,/\,
7.1\,\sim 9,000\,min^{-1}$  or $\rm 256^2\,/\, 21.2\,\sim 3,000\,min^{-1}$,
depending on  the readout interval, and about $\rm 25,000\,min^{-1}$
in record mode. These rates have never been exceeded since the DDS
reprogramming. 

The collection of data mostly relied on MROs when DDS detected 
relatively low impact and noise rates. The timing accuracy of these events, 
however, is less precise (TEV = 259). 

\begin{center} \fbox{\bf Insert Fig.~\ref{rate}} \end{center}

Table~2 lists the number of all dust impacts and noise events identified 
with the Galileo dust sensor in 1996 as deduced from the accumulators 
of classes 2 and 3. Depending on the event rate, the numbers
are given in intervals from half a day to a few weeks (the numbers with
the highest time resolution achievable are available in electronic form
only and are provided to the data archiving centers). For these two classes 
with the lowest amplitude range AR1 the complete data set for only 2\% of 
the detected events was transmitted, the 
remaining 98\% of events were only counted. Nearly all data sets for 
events in higher amplitude ranges were transmitted, although a few were also 
lost in AR2 and AR3. We give only the number 
of events in classes 2 and 3 because they have been shown to contain
real dust impacts: class~3 is practically noise free (although Kr\"uger et al.
 \cite*{krueger1999c} found indications for a very small number of noise 
events in class~3, AR1, in 
the inner Jovian system). Class~2 is strongly contaminated by noise events 
in the inner Jovian system (within about $\rm 20\,R_J$ from Jupiter). 

\begin{center} \fbox{\bf Insert Tab.~2} \end{center}

The data set we present here has been denoised according to the criteria 
derived by Kr\"uger et al. \cite*{krueger1999c}. The noise contamination factor 
$f_{noi}$ for class~2 listed in Tab.~2 for the six different amplitude 
ranges has been derived with two different methods: for AR1 we use the 
procedure described by \cite{krueger1999c}, i.e. a one day average of 
the ratio between the number of class~2 AR1 noise events and the total 
number of class~2 AR1 events (noise plus dust) for which the 
full data set is available. This defines the scaling factor $f_{noi}$ with which 
the number AC21 of events derived has to be 
multiplied in order to get the number of noise events from the 
AC21 accumulator. It should be noted that the criteria applied to identify
noise events in the DDS data have been derived with our present knowledge of 
the instrument behaviour in the Jovian environment. Future analyses may
lead to an improved picture of the noise characteristics and to modified 
algorithms for noise rejection.

For the higher amplitude ranges AR2 to AR6 the full 
data set is available for most detected events. Thus, from the charge amplitudes
and rise times we can 
determine for each accumulator increment listed in Tab.~2 whether it was 
due to a noise event or a dust impact. During time intervals when
the complete data set for some but not all detected events 
was transmitted, $f_{noi}$ has been calculated from those 
events for which the full data set is available and it has been assumed that
this noise ratio applies to all events detected (i.e. also counted) in this time 
interval. 

It should be noted that the noise identification criteria used here 
are exactly those derived in \cite{krueger1999c}. The analysis of the
latest data after 1997 has shown that the noise behaviour of the 
dust sensor has changed due to instrument ageing (see Section~\ref{analysis}).
Different noise identification criteria have to be applied 
to later data. Data from 1996
published in this paper are not affected. 

The noise identification criteria  of Kr\"uger et al. \cite*{krueger1999c}
which have been applied to the 1996 data have been developed to separate the 
tiny Jupiter stream particles from noise events. However, they do not work 
very well to distinguish secondary ejecta grains detected during close 
satellite fly-bys from noise events. We therefore 
applied a different technique to identify ejecta grains which is summarised in 
Tab.~\ref{sec_criteria} and \ref{denoise}. 

During all four satellite fly-bys in 
1996 the detection geometry was such that ejecta grains could be detected from 
rotation angles $\rm 180^{\circ} \leq ROT \leq 360^{\circ}$ only. 
During the fly-bys G1, G2 and E4 the impact direction (ROT) could be used as a 
good parameter to identify ejecta grains because stream particles and 
ejecta grains approached from opposite directions \cite{krueger1999d}. 
During the C3 fly-by, however, the stream particles approached from the same 
direction as the ejecta grains and the measured impact velocities of the dust 
particles had to be used as an additional parameter to identify the ejecta 
\cite{krueger2000}. The C3 fly-by velocity of Galileo was $\rm 8.0\, km\,s^{-1}$ 
and we included only particles with a measured impact velocity below 
$\rm 10.0\, km\,s^{-1}$ in the data set. For the Ganymede and Europa fly-bys 
we did not limit the velocity range of the grains. 

For all four satellite fly-bys of 1996 we included only particles within 
the approximate Hill radius of the satellite, except for Europa 
where we used a larger altitude limit because the present data analysis indicates 
that this dust cloud may be more extended.
Denoising has been shown to be important for class~2 in the inner Jovian 
system \cite{krueger1999c},  i.e. within about $\rm 15\,R_J$ from 
Jupiter. We therefore denoised the data from the Europa fly-by but did not
denoise the data from the Ganymede and Callisto fly-bys. For denoising of the
Europa data we used a slightly different noise separation scheme 
than Kr\"uger et al. \cite*{krueger1999c}. Events which fulfil the 
criteria listed in Tab.~\ref{denoise} have been rejected as noise.

\begin{center} \fbox{\bf Insert Tab.~\ref{sec_criteria}} \end{center}
\begin{center} \fbox{\bf Insert Tab.~\ref{denoise}} \end{center}

During the first three 
passages through the inner Jovian system (G1, G2 and C3) an 
unknown number of accumulator overflows occurred in the lowest amplitude range, 
especially in class~2. Therefore, the numbers before the instrument 
reprogramming on 4 December 1996 given in Tab.~2 should be treated as 
lower limits, specifically when the corresponding rates are close to the 
maximum recordable rates 
described above. For numbers after 13 December 1996 (when the instrument 
was read out in RTS mode) overflows of the
AC21 and AC31 accumulators are fully taken into account
and the numbers given in Tab.~2 represent the true numbers of detected 
events. Between 4 and 13 December accumulator overflows are also taken into 
account but due to high dust impact rate and the instrument readout mode with 
memory readouts occurring at a few day intervals unrecognized overflows 
may also have happened. Thus, in this time interval impact rates should also
be treated with caution.

To our present understanding the lower quality classes 0 and 1
contain only noise events and are therefore not considered here. Future
efforts, however, may also lead to the identification of some dust impacts 
in these low quality classes. 

The dust impact rate recorded by DDS in  1996 as deduced from the 
class 2 and 3 accumulators is shown in Fig.~\ref{rate}. The impact rate 
measured in the lowest amplitude range (AR1) and the one measured in the 
higher amplitude ranges (AR2 to AR6) are shown separately because they
reflect two distinct populations of dust.
AR1 contains mostly stream particles which have been
measured throughout the Jovian system. 
Bigger particles (AR2 to AR6) 
have been mostly detected between the Galilean
satellites.  This is illustrated in the diagram: the impact rate for AR1
gradually increases when Galileo approaches the 
inner Jovian system, whereas it shows narrow 
peaks close to the perijove passages in the case of the bigger 
(AR2 to AR6) impacts. 
%These bigger grains form  a dust ring between 
%the Galilean satellites which is maintained by particles that escape 
%from the clouds surrounding the Galileans \cite{krivov1999}. 
%The majority of them\cite{colwell1998,thiessenhusen2000}. The 
%prograde population is most likely maintained by grains escaping
%from the clouds surrounding the Galilean satellites \cite{krivov1999}
Diagrams showing the class~3 AR1 impact rate 
with a much higher time resolution and as a function of distance from 
Jupiter have been published by Gr\"un et al. \cite*{gruen1998} and are not repeated here.

The impact rates of AR1 particles measured in the inner Jovian system 
reached maximum values of 
about $\rm 20\,min^{-1}$ during the G1, G2, and C3 orbits. These values are 
close to the saturation limit caused by unrecognized accumulator overflows
(see above) and higher short-time peaks may have occurred. More than 100 impacts 
per minute have been detected in E4 (1996, day 350) which represents the highest 
impact rate recorded during Galileo's prime mission.
%$\rm 100\,min^{-1}$ in E4.
Such a high rate could only be recorded in RTS mode after the reprogramming on 
4 December 1996.

Table~5 lists the data sets for all 95 big particles detected in classes~2 and 3 between 
January and December 1996 for which the complete information exists.
Class~2 particles have been separated from noise by applying the criteria developed 
by Kr\"uger et al. \cite*{krueger1999c} except for the satellite fly-bys (see above).
We do not list the small stream particles (AR1) in Tab.~3 because their masses and 
velocities are outside the calibrated range of DDS and they are by far too 
abundant to be listed here (secondary ejecta grains in AR1 are also omitted). 
The complete information of a total of 5258 small
dust particles has been transmitted in 1996. The stream particles are 
believed to be about 10~nm in size and their velocities exceed 
$\rm 200~km\,s^{-1}$ \cite{zook1996}. Any masses and velocities derived 
for these particles with existing calibration algorithms would 
be unreliable. The full data set for all 5353 small and big particles is submitted 
to the data archiving centers and is available in electronic form. 
A total number of 9119 events (dust plus noise in all amplitude ranges and classes) 
were transmitted in 1996, each with a complete data set.

\begin{center} \fbox{\bf Insert Tab.~5} \end{center}

In Tab.~5 dust particles are identified by their sequence number 
and their impact time. Gaps in the sequence number are due to the 
omission of the small particles. The time error value (TEV) as 
defined above is listed next. Then the event category -- class (CLN) 
and amplitude range (AR) -- are given. Raw data as transmitted to Earth 
are displayed in the next columns: sector value (SEC) which is the
spacecraft spin orientation at the time of impact, 
impact charge numbers (IA, EA, CA) and rise times (IT, ET), time
difference and coincidence of electron and ion signals (EIT, EIC),
coincidence of ion and channeltron signal (IIC), charge reading at
the entrance grid (PA) and time (PET) between this signal and
the impact. Then the instrument configuration is given: event
definition (EVD), charge sensing thresholds (ICP, ECP, CCP, PCP) and
channeltron high voltage step (HV). See Paper~I for further
explanation of the instrument parameters, except TEV which is introduced
above.

The next four columns in Tab.~3 give information about Galileo's orbit: 
ecliptic longitude and latitude (LON, LAT) and distance from Jupiter ($\rm D_{Jup}$, 
in $\rm R_J$). The next column gives the 
rotation angle (ROT) as described in Sect.~\ref{mission}. 
Whenever this value is unknown, ROT is arbitrarily set to
999. This occurs 10 times in the full data set that includes 
the small particles. Then follows the pointing direction of DDS at 
the time of particle impact in ecliptic longitude and latitude 
($\rm S_{LON}$, $\rm S_{LAT}$).
When ROT is not valid $\rm S_{LON}$ and $\rm S_{LAT}$ are also useless. 
Mean
impact velocity (v) and velocity error factor (VEF, 
i.e. multiply or divide stated velocity by VEF to obtain upper or lower 
limits) as well as mean particle mass (m) and mass error factor (MEF) are 
given in the last columns. For VEF $> 6$, both velocity and mass values 
should be discarded. This occurs for 8 impacts. No intrinsic dust charge values 
are given \cite{svestka1996}.

Entries for the parameter PA in Tab.~3 sometimes have values between 49 and 63 
although the highest possible value is 48 (Paper~I). This is also 
inherent in all Galileo and Ulysses data sets published earlier (Papers~II to V) 
and it is due to a bit flip. According to our present understanding the correct PA 
values are obtained by subtracting 32 from all entries which have values between 
49 and 63. Values of 48 and lower should remain unchanged.

\section{Analysis} \label{analysis}

\begin{figure}
\epsfxsize=8.5cm
\epsfbox{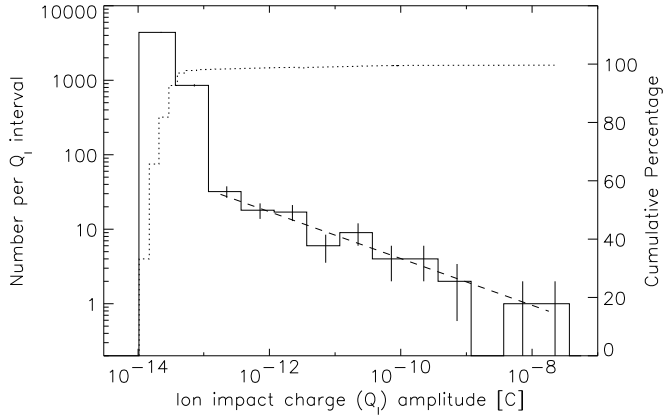}
%\epsfbox{../bilder/nqi.ps}
        \caption{\label{nqi}
Amplitude distribution of the impact charge $Q_I$ for the 5353 dust particles 
detected in 1996. The solid line
indicates the number of impacts per charge interval, whereas the 
dotted line shows the cumulative distribution. Vertical bars
indicate the $\rm \sqrt{n}$ statistical fluctuation. A power law fit
to the data for big particles with $Q_I >\rm  10^{-13}\,C$ (AR2 to AR6) 
is shown as a dashed line (power law index -0.31). 
}
\end{figure}

The positive charge measured on the ion collector, $Q_I$, is 
the most important impact parameter determined by DDS because it is 
rather insensitive to noise. Figure~\ref{nqi} shows the distribution of 
$Q_I$ for the full 1996 data set (small and big particles together).
Ion impact charges have been detected over the entire range of six 
orders of magnitude the instrument can measure. One impact (or about
0.02\% of the total) is close to the saturation limit of $ Q_I \sim 
\rm 10^{-8}\,C$ and may thus constitute a lower limit of the actual impact 
charge. The impact charge 
distribution of the big particles ($ Q_I >\rm 10^{-13}\,C$) follows a 
power law with index $-0.31$ and is shown as a dashed line. This slope
is close to the value of $-1/3$ given for Galileo in Paper~II for 
the inner solar system. It is flatter than the $-1/2$ given for Ulysses 
in Paper~III and the $-0.43$ 
given for Galileo in Paper~IV, which both mainly reflect the outer solar 
system. The slopes indicate that, on average, bigger particles 
have been detected in the inner solar system and in the Jovian system 
than in the interplanetary space of the outer solar system. This is in 
agreement with a smaller relative contribution of interstellar particles 
in the inner solar system and in the Jovian system. Note that the Jovian 
stream particles (AR1) have been excluded from the power law fit. 

\begin{center} \fbox{\bf Insert Fig.~\ref{nqi}} \end{center}

\begin{figure}
\vspace{-2cm}
\epsfxsize=8.5cm
\epsfbox{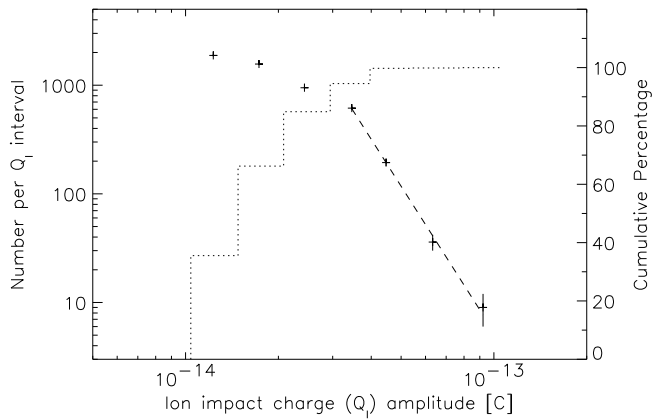}
%\epsfbox{../bilder/nqi2.ps}
        \caption{\label{nqi2}
Same as Fig.~\ref{nqi} but for the small particles in the lowest 
amplitude range (AR1) only. A power law fit to the data with 
$\rm 3\times 10^{-14}\,C < {\it Q_I} < 10^{-13}\, C$ is shown as a dashed 
line (power law index -4.46).
}
\end{figure}

In Fig.~\ref{nqi} the small stream particles ($ Q_I <\rm 10^{-13}\,C$) 
are collected in two histogram bins. Their number per individual 
digital step is shown separately in Fig.~\ref{nqi2} to analyse their 
behaviour in more detail. The distribution flattens 
for impact charges below $\rm 3\times 10^{-14}\,C$. This 
indicates that the sensitivity threshold of DDS may not be sharp and it 
is consistent with the number of impacts with the lowest impact charges 
$Q_I$ not being complete due to the low data transmission capability of 
Galileo. The impact charge distribution for small particles 
with $Q_I >\rm  3\times 10^{-14}\,C$ follows a power law with index 
-4.5. This indicates that the size distribution of the small stream 
particles rises steeply towards smaller particles. It is much 
steeper than the distribution of the big particles shown in Fig.~\ref{nqi}. 

\begin{center} \fbox{\bf Insert Fig.~\ref{nqi2}} \end{center}

\begin{figure}
\vspace{-1cm}
\epsfxsize=8.5cm
\epsfbox{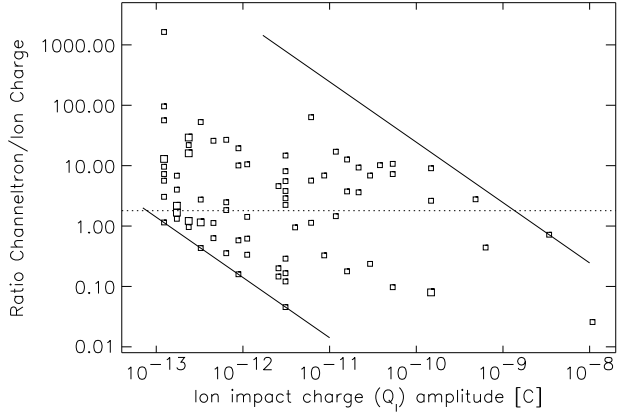}
%\epsfbox{../bilder/qiqc.ps}
        \caption{\label{qiqc}
Channeltron amplification factor $A = Q_C/Q_I$
as a function of impact charge $Q_I$ for big particles (AR2 to AR6)
detected in 1996. The solid lines indicate the sensitivity
threshold (lower left) and the saturation limit (upper right) of the channeltron. 
Squares indicate dust particle impacts, and the area of the squares is proportional 
to the number of events (the scaling of the squares is not the same as in
earlier papers). The dotted horizontal line shows the mean value 
of the channeltron amplification A\,=\,1.8 for ion impact charges 
$\rm 10^{-12}~C < {\it Q_I} < 10^{-10}~C$.
}
\end{figure}

The ratio of the channeltron charge $Q_C$ and the ion collector
charge $Q_I$ is a measure of the channeltron amplification A which
is an important parameter for dust impact identification (Paper~I).
The in-flight channeltron amplification was determined 
in Papers~II and IV for the initial six years of the 
Galileo mission to identify a possible degrading of the 
channeltron. For a channeltron high voltage of 1020~V (HV~=~2) 
the amplification $Q_C/Q_I$ obtained for $\rm 
10^{-12}{\rm\, C} \le {\it Q_I} \le 10^{-10}{\rm\, C}$ was $\rm A \sim 1.6$
and $\rm \sim 1.4$, respectively.
Here we repeat the same analysis for the 1996 data set.
Figure~\ref{qiqc}
shows the charge ratio $Q_C/Q_I$ as a function of 
$Q_I$ for the same high voltage as in the previous papers. The charge 
ratio $Q_C/Q_I$ determined for $\rm 10^{-12}{\rm\,C} \le {\it Q_I }
\le 10^{-10}{\rm\,C}$ is $\rm A \sim 1.8$. This is close to 
the earlier values and shows that there is no ageing 
of the channeltron detectable. Channeltron ageing is seen in the 
data after 1996 which will be the subject of a future publication.

\begin{center} \fbox{\bf Insert Fig.~\ref{qiqc}} \end{center}

\begin{figure}
\vspace{-1cm}
\epsfxsize=8.5cm
\epsfbox{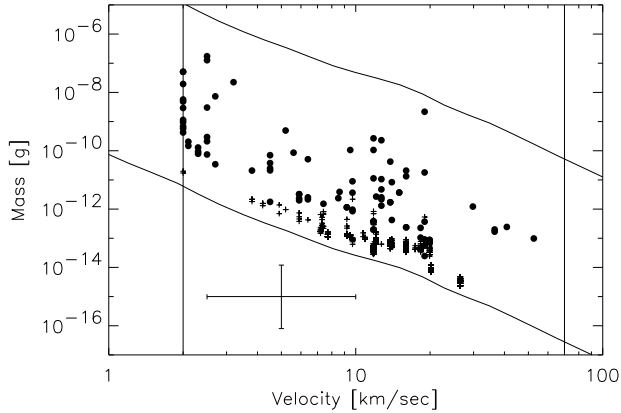}
%\epsfbox{../bilder/mass_speed.ps}
        \caption{\label{mass_speed}
Masses and impact speeds of all 5353 impacts recorded by DDS in 1996.
The lower and upper solid lines indicate the threshold and
saturation limits of the detector, respectively, and the vertical lines 
indicate the calibrated velocity range. A sample error bar is shown that indicates
a factor of 2 error for the velocity and a factor of 10 for the mass determination.
Note that the small particles (plus signs) are probably faster and smaller 
than implied by this diagram (see text for details).
}
\end{figure}

Figure~\ref{mass_speed} displays the calibrated masses and 
velocities of all 5353 dust grains detected in 1996. 
Impact velocities have been measured over almost the entire 
calibrated range from 2 to $\rm 70\,km\,s^{-1}$, and 
the masses vary over 8 orders of magnitude from $\rm 10^{-7}\,g$ to 
$\rm 10^{-15}\,g$. The mean errors are a factor of 2 for the 
velocity and a factor of 10 for the mass. Impact 
velocities below about $\rm 3\,km\,s^{-1}$ should be treated with caution.
Anomalous impacts onto the sensor grids or structures 
other than the target generally lead to prolonged rise times of the 
charge signals. This in turn results in artificially low impact 
velocities and high dust particles masses. 

\begin{center} \fbox{\bf Insert Fig.~\ref{mass_speed}} \end{center}

The mass range populated by the particles is by 
two orders of magnitude smaller than that reported from the initial 
six years of the mission. The largest and 
smallest masses reported earlier, however, are at the edges of the 
calibrated velocity range of DDS and, hence, they 
are the most uncertain. 
Any clustering of the velocity values is due to discrete steps in 
the rise time measurement but this quantization is much smaller than the
velocity uncertainty. Masses and velocities in the lowest 
amplitude range (AR1, particles indicated by plus signs) should be 
treated with caution. These are mostly Jovian stream 
particles for which we have clear indications that their masses 
and velocities are outside the calibrated range of DDS \cite{zook1996}
(J. C. Liou, priv. comm.). The particles are probably 
much faster and smaller 
than implied by Fig.~\ref{mass_speed}. On the other hand, the 
mass and velocity calibration is valid for the bigger particles
\cite{krueger1999d,krueger2000}. 
For many particles in the lowest two amplitude ranges (AR1 and 
AR2) the velocity had to be computed from the ion charge signal 
alone which leads to the striping in the lower mass range in 
Fig.~\ref{mass_speed} (most 
prominent above $\rm 10\,km\,s^{-1}$). In the 
higher amplitude ranges the velocity could normally be calculated 
from both the target and the ion charge signal which leads to  
a more continuous distribution in the mass-velocity plane. 

Although no ageing of the channeltron could be found from 
Fig.~\ref{qiqc} with the 1996 data set, other indications for 
ageing of DDS caused by 
the effects of the harsh radiation environment in the inner 
Jovian magnetosphere have been found: (1) The measured instrument
current, which had a constant value of about 77.5 mA during the interplanetary cruise 
of Galileo, began to drop by 3\% per year when the spacecraft
was injected into the Jovian system in December 1995. 
This drop is mostly likely caused by radiation-induced 
ageing of a resistor and is inherent to the measurement process 
itself rather than being related to a real drop in the instrument
current. (2) Changes in the test pulses generated by 
the instrument-built in test 
pulse generator. (3) A drift in the mean target and ion 
collector rise time signals ET and IT. Although it is best 
recognized in the data set for AR1, it may also affect the higher
amplitude ranges. This drift does not affect the calibration of
the 1996 data set but it may have to be taken into account in the mass 
and speed calibration of later data. The consequences of these ageing effects 
are under investigation and will be the subject of a future paper.

\section{Discussion} \label{discussion}

\begin{figure}[ht]
\vspace{-5cm}
\epsfxsize=16cm
\epsfbox{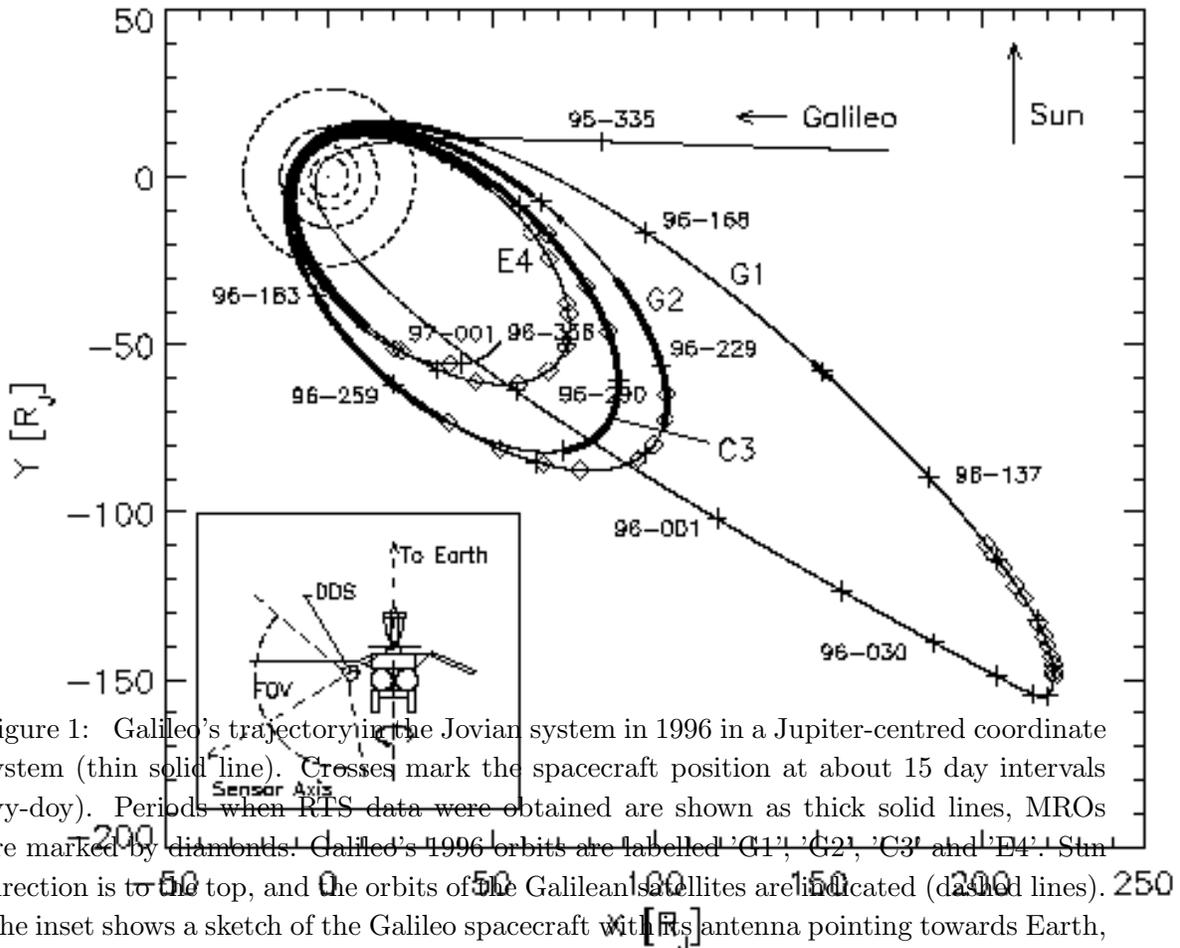}
%\epsfbox{../bilder/trajectory.ps}
\vspace{-4cm}
        \caption{\label{trajectory}
Galileo's trajectory in the Jovian system in 1996 in a Jupiter-centred 
coordinate system (thin solid line). 
Crosses mark the spacecraft position at about 15 day intervals (yy-doy).
Periods when RTS data were obtained are shown as thick solid lines,
MROs are marked by diamonds. Galileo's 1996 orbits are labelled 
'G1', 'G2', 'C3' and 'E4'. Sun direction is to the top, and 
the orbits of the Galilean satellites are indicated  (dashed lines).
The inset shows a sketch of the Galileo spacecraft with its 
antenna pointing towards Earth, the dust detector (DDS) and its 
field of view (FOV). DDS makes about 3 revolutions per minute.
Sun and Earth direction coincide to within $10^{\circ}$.
}
\end{figure}

\begin{figure}
\epsfxsize=15cm
\epsfbox{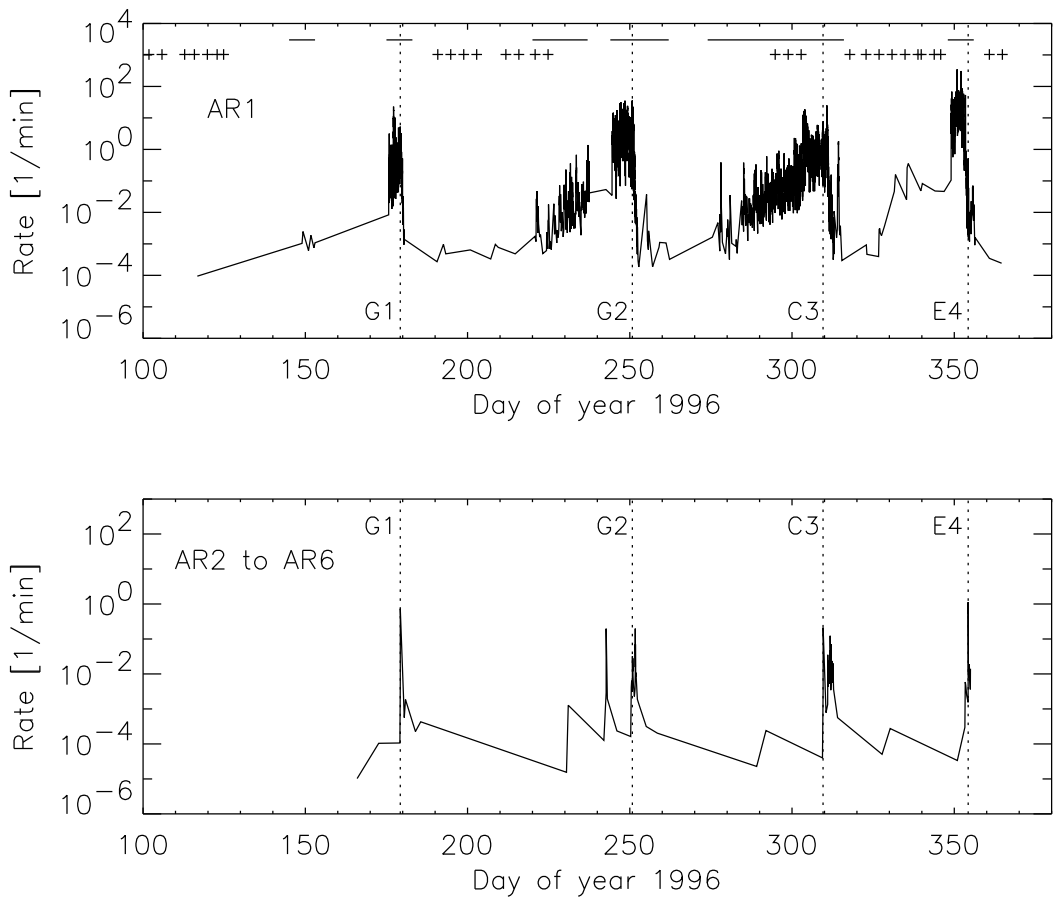}
%\epsfbox{../bilder/rate.ps}
        \caption{\label{rate}
Dust impact rate detected by DDS in 1996. The upper panel shows the 
impact rate in AR1 which represents the Io 
dust streams, the lower panel that for the  higher amplitude ranges AR2 
to AR6. Dotted lines indicate the closest approaches to the Galilean satellites.
Perijove passages occurred within two days of the satellite closest approaches.
These curves are plotted from the number of impacts with the highest time resolution
which is available only in electronic form. No smoothing has been applied which leads
to the 'sawtooth' pattern, especially prominent in the lower panel. In the upper 
panel, time intervals with continuous RTS data coverage are indicated by 
horizontal bars, memory readouts (MROs) are marked by crosses.
}
\end{figure}

\begin{figure}
\epsfxsize=14.5cm
\epsfbox{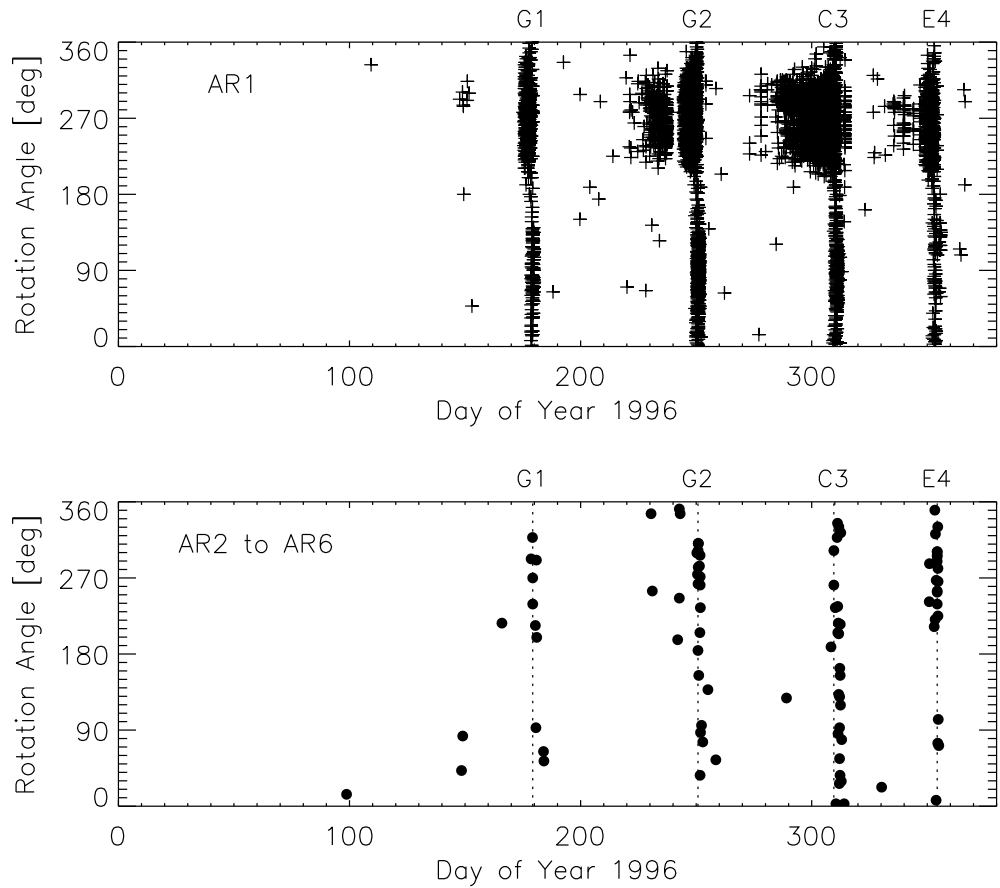}
%\epsfbox{../bilder/rot_angle.ps}
        \caption{\label{rot_angle} 
Rotation angle vs. time for two different mass ranges,
upper panel: small particles, AR1 (Io dust stream particles);
lower panel: big particles, AR2 to AR6. See 
Sect.~\ref{mission} for an explanation of the rotation angle.
The encounters with the Galilean satellites are indicated
by dashed vertical lines.
}
\end{figure}

By far the largest number of particles in the 1996 data set presented 
here are tiny dust grains originating from Io 
\cite{horanyi1997,gruen1998,graps2000,heck1998}.
These grains almost exclusively populate amplitude range AR1
(see also Fig.~\ref{rate}, upper panel). They approach the sensor 
as collimated streams and their 
impact direction shows a characteristic behaviour that can only
be explained by grains having a radius of about 10\,nm and which
strongly interact with the Jovian magnetosphere 
\cite{gruen1998}. 
%The strong magnetospheric 
%interaction is also demonstrated by clear 5 and 10\,h periodicities 
%in the impact rate. 
The impact direction of these grains is shown in 
the upper panel of Fig.~\ref{rot_angle}.
On the inbound trajectory, when Galileo approaches Jupiter, the grains
were mainly detected from rotaton angles $270 \pm 70^{\circ}$. This
is best seen in the G2 and C3 orbits (1996, days 220 to 250 and 275 to 310) 
when we had continuous data coverage for the longest time period. One to 
two days before perijove passage the impact direction shifted by 
$180^{\circ}$ and the particles 
approached from $90 \pm 70^{\circ}$. Rotation angles of $90^{\circ}$ and 
$270^{\circ}$ are close to the ecliptic plane. The detection geometry is 
also seen in Fig.~\ref{trajectory}.

\begin{center} \fbox{\bf Insert Fig.~\ref{rot_angle}} \end{center}

The tiny dust stream particles could be used for an analysis of 
the sensitive area of the Galileo dust sensor \cite{krueger1999c} 
which could not be done during ground calibration. 
A detailed analysis of the distribution of the measured 
rotation angles showed that three of the other instruments on board Galileo 
obscure the field of view of the dust sensor. This can be seen in 
the upper panel of Fig.~\ref{rot_angle} (see also Kr\"uger et al.,
\cite*{krueger1999c}: there is a reduced number of 
particles with rotation angles $\rm ROT = 270 \pm 20^{\circ}$ in the G2 and 
C3 orbits when Galileo approaches Jupiter. This is best seen on days 200 to 225 
and 270 to 285 in 1996. Fewer particles were detected in this time and rotation angle
range with respect to the range $\rm ROT= 310 \pm 20^{\circ}$ and 
$\rm ROT=230 \pm 20^{\circ}$. This is caused by obscuration of the field 
of view. 

Furthermore, the times of the onset of the dust impacts measured in classes 
2 and 3 differ significantly which indicated that different sensitive areas 
apply to stream particles detected as class~2 or 3 impacts, respectively.
The times of the onset, $180^{\circ}$-shift and cessation of the dust streams 
are given in Tab.~\ref{stream_table}. Reliable onset times could be 
determined for class~3 and for orbits G2 and C3 only. For class~2 and for 
the other orbits no RTS data were obtained at the times of onset of 
the dust impacts. Note that Kr\"uger et al. \cite*{krueger1999c} used a larger 
data set for their analysis of the sensitive area of DDS which included 
data from 1997. 

During the $180^{\circ}$-shift  the detection rate of class~3 dropped 
significantly whereas no such drop was seen in the class~2 impact rate. 
This is consistent with a reduced field of view for class~3 w.r.t. class~2
and the reader is referred to Kr\"uger et al. \cite*{krueger1999c} for details.
Especially, the shift for classes~2 and 3 occurred at the same time. Similarly,
no time difference in the cessation of impacts in the two classes is noticable
because at the cessation the dust streams sweep through the field of view rather 
quickly.

\begin{center} \fbox{\bf Insert Tab.~\ref{stream_table}} \end{center}

\begin{figure}
\epsfxsize=14.6cm
\epsfbox{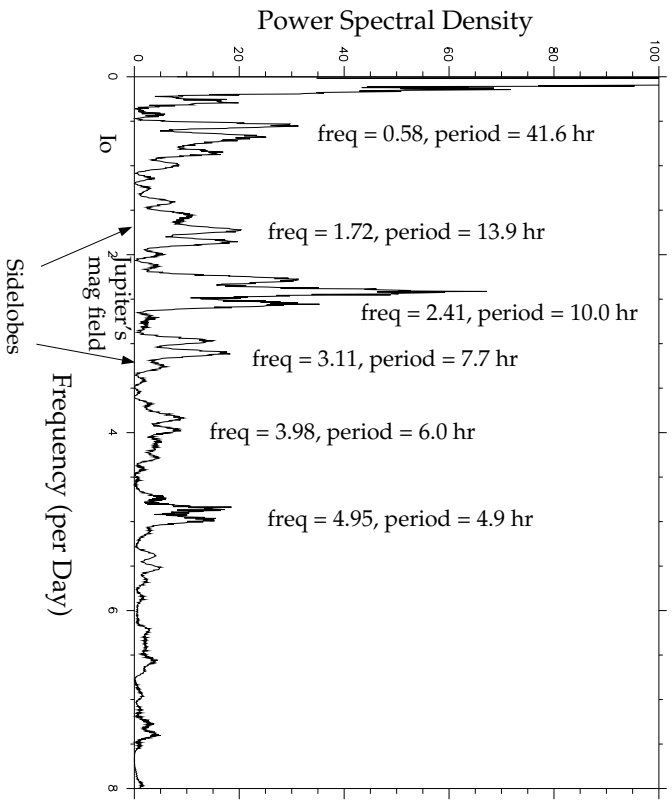}
%\epsfbox{../bilder/periodo.ps}
\vspace{-6.5cm}
\caption{\label{periodogram}
A periodogram 
for the dust impact rate detected
in 1996. See text for details.
}
\end{figure}

A periodogram \cite{scargle1982}    
of the dust impact rate measured in 1996 shows 5 and 10\,h
periodicities which are caused by the interaction of the dust grains with Jupiter's
rotating magnetosphere (Fig.~\ref{periodogram}). Another strong peak occurs at 
Io's orbital frequency of 42\,h. In addition, there are side lobes at 
Io's orbital frequency plus or minus Jupiter's rotation frequency (10\,h
rotation period) or twice that frequency (5\,h), respectively. 
If Io's orbital frequency is a carrier frequency, then the side frequencies 
show that Jupiter's rotation frequency amplitude-modulates Io's signal.
Graps et al. \cite*{graps2000} have analysed a larger data set from 1996 
and 1997 and used these findings -- among other arguments -- to conclude that 
Io is the source of the dust streams \cite{graps2000}.

\begin{center} \fbox{\bf Insert Fig.~\ref{periodogram}} \end{center}

The lower panel of Fig.~\ref{rot_angle} shows the rotation angle for 
a second population of dust grains, namely bigger micrometer-sized grains.
These grains are concentrated in the inner Jovian system forming a 
tenuous dust ring between the Galilean satellites. Modelling \cite{krivov2001} 
has shown that this ring is fed by particles escaping from impact-generated 
dust clouds around the Galilean moons \cite{krueger1999d,krueger2000}. 
The impact directions and impact times imply that
two groups of grains exist in the dust ring: particles 
on prograde and retrograde orbits
about Jupiter, respectively \cite{colwell1998,thiessenhusen2000}. 
Prograde particles are much more
abundant than retrograde ones \cite{thiessenhusen2000}. A fraction 
of the retrograde grains may be interplanetary or interstellar particles 
captured by the Jovian magnetosphere \cite{colwell1998}.

\hspace{1cm}

{\bf Acknowledgements.}
The authors thank the Galileo project at JPL for effective and successful 
mission operations. We are grateful to Mark Sykes whose careful evaluations
improved the Galileo and Ulysses dust data sets submitted
to the Planetary Data System. We thank our referees, Alexander V. Krivov and
Larry W. Esposito, for providing valuable suggestions which improved the
presentation of our results. This work has been supported by the Deutsches
Zentrum f\"ur Luft- und Raumfahrt (DLR). 

\renewcommand{\baselinestretch}{1.1} % Zeilenabstand 1.1 * Defaultwert

\bibliography{/home/krueger/tex/bib/pape,/home/krueger/tex/bib/references}

%\vspace{2cm}
%\begin{itemize}
%\item[] Ejekta-Teilchen (Tabelle), St\"orungen bei den Vorbeifl\"ugen
%\item[] Power in Frequenzen der einzelnen Orbits
%\end{itemize}

\clearpage

%\section{Tables}
%\setlength{\parskip}{4pt plus 1pt minus 1 pt}   %etwas gr"o"sere Abst"ande
%\setlength{\parindent}{0pt}                     %Keine Absatzeinrueckung
%

\thispagestyle{empty}
\begin{table}[htb]
\caption{\label{event_table} Galileo mission and dust detector (DDS) 
configuration, tests and other events. See text for details. Only 
selected events are given before 1996. Distances from Jupiter 
are measured from the center of the planet, altitudes are measured 
from the surface (Jupiter radius $R_J =71492\,\rm km$).
}
\scriptsize
  \begin{tabular*}{\hsize}{lccl}
   \hline
   \hline \\[-2.0ex]
%Titelzeile
Yr-day& 
Date&
Time& 
Event \\[0.7ex]
\hline \\[-2.0ex]
89-291& 18 Oct 1989& 16:52& Galileo launch \\
95-341& 07 Dec 1995& 21:54& Galileo Jupiter closest approach, distance: $\rm 4.0\,R_J$  \\
95-341& 07 Dec 1995& 23:25& DDS configuration: HV=off \\
96-087& 27 Mar 1996& 05:56& DDS configuration: HV=2, EVD =C,I, SSEN = 0,0,1,1 \\
96-089& 29 Mar 1996& 16:34& DDS first MRO after high voltage switch on \\
96    &April/May 1996&    & Galileo reprogramming (phase 2 software) \\
96-145& 24 May 1996& 21:00& DDS begin RTS data after Galileo reprogramming\\
96-153& 01 Jun 1996& 10:15& DDS end RTS data\\
96-175& 23 Jun 1996& 16:17& DDS begin RTS data\\ %, 3\,bps \\
96-175& 23 Jun 1996& 17:04& Galileo turn: $20^{\circ}$, turn to nominal attitude \\
96-176& 24 Jun 1996& 18:30& Galileo OTM-6, duration 5\,h, no attitude change \\
96-179& 27 Jun 1996& 06:07& DDS end RTS data, begin record data \\ %, 3\,bps \\
96-179& 27 Jun 1996& 06:29& {\bf Galileo Ganymede 1 (G1) closest approach}, altitude 835\,km \\
96-179& 27 Jun 1996& 06:53& DDS end record data, begin RTS data \\ 
96-179& 27 Jun 1996& $\rm \sim 12\,h$ & DDS begin deadtime caused by strong channeltron noise \\
96-180& 28 Jun 1996& 00:31& Galileo Jupiter closest approach, distance $\rm 11\,R_J$ \\
96-180& 28 Jun 1996& 22:48& Galileo turn: $33.4^{\circ}$, duration 9.1\,h \\
96-181& 29 Jun 1999& $\rm \sim 0\,h$ & DDS end deadtime caused by strong channeltron noise \\ 
96-181& 29 Jun 1996& 06:56& Galileo turn: $28.4^{\circ}$, duration 42\,h \\
96-182& 30 Jun 1996& 07:45& Galileo OTM-7A, duration 15\,h, no attitude change \\
96-183& 01 Jul 1996& 01:03& Galileo turn: $5.1^{\circ}$, return to nominal attitude \\
96-183& 01 Jul 1996& 08:52& DDS end RTS data \\ 
96-185& 03 Jul 1996& 09:00& Galileo OTM-7B, duration 15\,h, no attitude change \\
96-218& 05 Aug 1996& 08:00& Galileo OTM-8,  duration 8\,h, size of turn $15^{\circ}$ \\
96-220& 07 Aug 1996& 03:28& DDS begin RTS data \\
96-232& 19 Aug 1996& 02:18& Galileo turn: $3.8^{\circ}$, return to nominal attitude \\
96-237& 24 Aug 1996& 14:15& Galileo spacecraft anomaly, end of RTS data \\
96-240& 27 Aug 1996& 17:30& Galileo OTM-9, duration 5\,h, no attitude change \\
96-244& 31 Aug 1996& 12:30& DDS begin RTS data after spacecraft anomaly \\
96-248& 04 Sep 1996& 18:50& Galileo OTM-10, duration 4.8\,h, no attitude change \\
96-250& 06 Sep 1996& 11:20& DDS configuration: HV=2, EVD =I, SSEN = 0,1,1,1, $\rm 18\,R_J$ from Jupiter \\
96-250& 06 Sep 1996& 18:32& DDS end RTS data, begin record data \\ 
96-250& 06 Sep 1996& 19:00& {\bf Galileo Ganymede 2 (G2) closest approach}, altitude 262\,km\\
96-250& 06 Sep 1996& 19:28& DDS end record data, begin RTS data \\ 
96-250& 06 Sep 1996& 20:07& Galileo turn: $23^{\circ}$, duration 3\,h, return to nominal attitude  \\ 
96-251& 07 Sep 1996& 13:38& Galileo Jupiter closest approach, distance $\rm 10.7\,R_J$ \\ 
96-252& 08 Sep 1996& 16:07& DDS configuration: HV=2, EVD =C,I, SSEN = 0,0,1,1, $\rm 18\,R_J$ from Jupiter \\
96-253& 09 Sep 1996& 21:30& Galileo OTM-11, duration 8\,h, size of turn $88^{\circ}$, return to nominal attitude  \\
96-255& 11 Sep 1996& 02:39& DDS end RTS data, begin record data \\ 
96-255& 11 Sep 1996& 03:19& DDS end record data, begin RTS data \\ 
96-262& 18 Sep 1996& 16:41& DDS end RTS data \\ 
96-267& 23 Sep 1996& 17:00& Galileo turn: $5^{\circ}$, new nominal attitude \\
96-274& 30 Sep 1996& 23:08& DDS begin RTS  data\\ 
96-282& 08 Oct 1996& 14:30& Galileo OTM-12, duration 7.5\,h, no attitude change \\
96-306& 01 Nov 1996& 13:30& Galileo OTM-13, duration 5\,h, no attitude change \\
96-309& 04 Nov 1996& 13:15& DDS end RTS data, begin record data \\ 
96-309& 04 Nov 1996& 13:34& {\bf Galileo Callisto 3 (C3) closest approach}, altitude 1,136\,km\\
96-309& 04 Nov 1996& 14:01& DDS end record data, begin RTS data \\ 
96-310& 05 Nov 1996& 10:11& DDS configuration: HV=2, EVD =I, SSEN = 0,1,1,1, $\rm 18\,R_J$ from Jupiter \\
96-311& 06 Nov 1996& 13:31& Galileo Jupiter closest approach, distance $\rm 9.2\,R_J$ \\
96-312& 07 Nov 1996& 03:10& Galileo turn: $16.2^{\circ}$, duration 17.5\,h, return to nominal attitude \\
\hline
\end{tabular*}\\[1.5ex]
\end{table}

\setcounter{table}{0}
\begin{table}[htb]
\caption{\label{event_table2} continued.
}
\scriptsize
  \begin{tabular*}{\hsize}{lccl}
   \hline
   \hline \\[-2.0ex]
%Titelzeile
Yr-day& 
Date&
Time& 
Event \\[0.7ex]
\hline \\[-2.0ex]
96-312& 07 Nov 1996& 16:42& DDS configuration: HV=2, EVD =C,I, SSEN = 0,0,1,1, $\rm 18\,R_J$ from Jupiter \\
96-313& 08 Nov 1996& 23:20& Galileo turn: $98.8^{\circ}$, duration 11\,h, return to nominal attitude \\
96-315& 10 Nov 1996& 07:20& Galileo OTM-14, no attitude change \\
96-316& 11 Nov 1996& 02:01& DDS end RTS data \\
96-316& 11 Nov 1996& 20:00& Galileo turn: $5.9^{\circ}$, new nominal attitude \\
96-331& 26 Nov 1996& 11:50& Galileo OTM-15, no attitude change \\
96-339& 04 Dec 1996& 16:33& DDS last MRO before reprogramming \\
96-339& 04 Dec 1996& 17:00& DDS reprogramming (AC21 and AC31 overflow counters added) \\
96-340& 05 Dec 1996& 03:50& DDS first MRO after reprogramming \\
96-348& 13 Dec 1996& 19:15& DDS begin RTS  data\\
96-351& 16 Dec 1996& 02:30& Galileo OTM-16, duration 5\,h, no attitude change \\
96-353& 18 Dec 1996& 11:00& DDS configuration: HV=2, EVD =I, SSEN = 0,1,1,1, $\rm 18\,R_J$ from Jupiter \\
96-354& 19 Dec 1996& 03:22& Galileo Jupiter closest approach, distance $\rm 9.2\,R_J$ \\ 
96-354& 19 Dec 1996& 06:26& DDS end RTS data, begin record data \\
96-354& 19 Dec 1996& 06:53& {\bf Galileo Europa 4 (E4) closest approach}, altitude 692\,km \\
96-354& 19 Dec 1996& 07:24& DDS end record data, begin RTS data \\
96-355& 20 Dec 1996& 04:00& Galileo turn: $80.6^{\circ}$, duration 14.4\,h, return to nominal attitude  \\
96-355& 20 Dec 1996& 06:43& DDS configuration: HV=2, EVD =C,I, SSEN = 0,0,1,1, $\rm 18\,R_J$ from Jupiter \\
96-356& 21 Dec 1996& 19:00& DDS end RTS data \\
96-358& 23 Dec 1996& 09:30& Galileo OTM-17, duration 9\,h, no attitude change \\
96-361& 26 Dec 1996& 20:38& Galileo turn: $11.0^{\circ}$, new nominal attitude \\           
\hline
\end{tabular*}\\[1.5ex]
Abbreviations used: MRO: DDS
memory read-out; HV: channeltron high voltage step; EVD: event definition,
ion- (I), channeltron- (C), or electron-channel (E); SSEN: detection thresholds,
ICP, CCP, ECP and PCP; OTM: orbit trim maneuver; RTS: Realtime science; 
AC21: class~2 amplitude range~1 accumulator; AC31: class~3 amplitude range~1 accumulator
\end{table}

\clearpage

\pagestyle{empty}

\begin{sidewaystable}
\tabcolsep1.5mm
\tiny
\vbox{
\hspace{-2cm}
\begin{minipage}[t]{22cm}
{\bf Table 2:} Overview of dust impacts accumulated with Galileo DDS 
between 1 January and 31 December 1996. 
The Jovicentric distance $D_{Jup}$, 
the lengths of the time interval $\Delta $t (days) from the previous 
table entry, and the corresponding numbers of impacts are given for the 
class 2 and 3 accumulators. The accumulators are arranged with increasing signal 
amplitude ranges (AR), 
e.g.~AC31 means counter for CLN = 3 and AR = 1. The determination of the 
noise contamination $f_{noi}$ in class~2 is described in the text. 
The $\Delta $t in the first line (day 96-148) 
is the time interval counted from the last entry in Table~4 in 
Paper~IV. The DDS reprogramming on day 96-339 is indicated by horizontal lines. 
The totals of  counted impacts, of impacts with 
complete data, and of all events (noise plus impact events) 
for the entire period are given as well.
\end{minipage}
}
\bigskip
\hspace{-2cm}
\begin{tabular}{|r|r|r|r|ccc|ccc|ccc|ccc|ccc|ccc|}
\hline
&&&& &&&& &&&& &&&& &&&& &\\
\mc{1}{|c|}{Date}&
\mc{1}{|c|}{Time}&
\mc{1}{|c|}{$D_{Jup}$}&
\mc{1}{|c|}{$\Delta $t}&
%\mc{4}{|c|}{AR-1}&
%\mc{4}{|c|}{AR-2}&
%\mc{4}{|c|}{AR-3}&
%\mc{4}{|c|}{AR-4}&
%\mc{4}{|c|}{AR-5}&
%\mc{4}{|c|}{AR-6}\\
{\scriptsize $f_{noi,AC21}$}&{\scriptsize AC}&{\scriptsize AC}&{\scriptsize $f_{noi,AC22}$}&
{\scriptsize AC}&{\scriptsize AC}&{\scriptsize $ f_{noi,AC23}$}&{\scriptsize AC}&
{\scriptsize AC}&{\scriptsize $f_{noi,AC24}$}&{\scriptsize AC}&{\scriptsize AC}&
{\scriptsize $ f_{noi,AC25}$}&{\scriptsize AC}&{\scriptsize AC}&{\scriptsize $ f_{noi,AC26}$}&
{\scriptsize AC}&{\scriptsize AC}\\
&&[$\rm R_J$]&\mc{1}{c|}{[d]}&
&$21^{\ast}$&$31^{\ast}$&
&22&32&
&23&33&
&24&34&
&25&35&
&26&36\\
&&&& &&&& &&&& &&&& &&&& &\\
\hline
&&&& &&&& &&&& &&&& &&&& &\\
96-148&11:14&      175.5& 151.9& 0.80&5&-&-&-&-&-&-&1&-&-&-&-&-&-&-&-&-\\
96-165&23:03&      107.9& 17.49& 0.27&11&-& 0.00&1&1&-&-&-&-&-&-&-&-&-&-&-&-\\
96-176&00:03&      45.10& 10.04& 0.17&134&269& 0.00&1&-&-&-&-&-&-&-&-&-&-&-&-&-\\
96-177&00:50&      36.22& 1.032& 0.00&304&64&-&-&-&-&-&-&-&-&-&-&-&-&-&-&-\\
96-178&14:13&      21.40& 1.558& 0.02&4133&938& 1.00&1&-&-&-&-&-&-&-&-&-&-&-&-&-\\
 &&&& &&&& &&&& &&&& &&&& &\\
96-179&03:19&      16.18& 0.545& 0.12&972&31& 0.00&1&-&-&-&-&-&-&-&-&-&-&-&-&-\\
96-180&10:14&      12.38& 1.288& 0.30&1398&29& 0.50&2&2&-&-&-&-&-&-&-&-&-&-&-&-\\
96-181&00:16&      17.14& 0.584& 1.00&37&-& 1.00&$\,$3$^{\ast}$ &-& 0.00&1&-&-&-&-&-&-&-&-&-&-\\
96-182&00:20&      26.55& 1.002& 1.00&3&-&-&-&-&-&-&-&-&-&-&-&-&-&-&-&-\\
96-185&13:18&      53.60& 3.540& 1.00&1&-& 0.50&2&-&-&-&-&-&-&1&-&-&-&-&-&-\\
 &&&& &&&& &&&& &&&& &&&& &\\
96-194&04:23&      94.16& 8.628& 0.74&15&-& 1.00&2&-&-&-&-&-&-&-&-&-&-&-&-&-\\
96-209&11:05&      125.3& 15.27& 0.74&17&-& 1.00&1&-&-&-&-&-&-&-&-&-&-&-&-&-\\
96-221&14:51&      125.7& 12.15& 0.77&30&-& 1.00&1&-&-&-&-&-&-&-&-&-&-&-&-&-\\
96-231&00:38&      111.9& 9.407& 0.32&91&5& 1.00&1&-&-&-&2&-&-&-&-&-&-&-&-&-\\
96-241&23:55&      75.75& 10.97& 0.07&444&73& 1.00&$\,$25$^{\ast}$ &1&-&-&-&-&-&-&-&-&-&-&-&-\\
 &&&& &&&& &&&& &&&& &&&& &\\
96-242&17:05&      72.30& 0.715& 0.12&246&173& 0.67&$\,$3$^{\ast}$ &-&-&-&-&-&-&-&-&-&-&-&-&-\\
96-245&00:11&      59.93& 2.295& 0.00&2595&666& 0.00&2&-&-&-&-&-&-&-&-&-&-&-&-&-\\
96-246&00:36&      53.67& 1.017& 0.04&5091&1202& 1.00&1&-&-&-&-&-&-&-&-&-&-&-&-&-\\
96-247&00:05&      47.11& 0.978& 0.03&4436&1139&-&-&-&-&-&-&-&-&-&-&-&-&-&-&-\\
96-247&21:19&      40.62& 0.884& 0.03&4897&1392& 1.00&1&-&-&-&-&-&-&-&-&-&-&-&-&-\\
 &&&& &&&& &&&& &&&& &&&& &\\
96-248&00:01&      39.75& 0.113& 0.00&765&168&-&-&-&-&-&-&-&-&-&-&-&-&-&-&-\\
96-249&00:10&      31.52& 1.005& 0.00&8443&2358&-&-&-&-&-&-&-&-&-&-&-&-&-&-&-\\
96-250&06:53&      19.74& 1.279& 0.05&8849&1213&-&-&1&-&-&-&-&-&-&-&-&-&-&-&-\\
96-251&02:47&      12.38& 0.829& 0.25&6314&400& 0.75&4&1&-&-&2& 0.00&1&1&-&-&1&-&-&-\\
96-252&07:27&      14.69& 1.194& 0.50&891&108& 0.14&7&-& 0.00&1&1&-&-&1&-&-&-&-&-&-\\
 &&&& &&&& &&&& &&&& &&&& &\\
96-255&01:49&      38.94& 2.765& 0.97&14&-& 0.67&3&-&-&-&-&-&-&2&-&-&-&-&-&-\\
96-260&19:33&      72.31& 5.739& 0.97&77&83& 1.00&1&-&-&-&-&-&-&1&-&-&-&-&-&-\\
96-270&22:28&      103.2& 10.12& 0.67&3&-& 1.00&1&-&-&-&-&-&-&-&-&-&-&-&-&-\\
96-285&16:03&      111.5& 14.73& 0.25&147&1& 1.00&1&-&-&-&-&-&-&-&-&-&-&-&-&-\\
96-291&23:09&      103.6& 6.296& 0.23&194&2& 1.00&1&1&-&-&-&-&-&-&-&-&-&-&-&-\\
 &&&& &&&& &&&& &&&& &&&& &\\
96-305&00:16&      58.77& 13.04& 0.01&8485&3324&-&-&-&-&-&-&-&-&-&-&-&-&-&-&-\\
96-306&00:03&      52.93& 0.991& 0.07&1607&184&-&-&-&-&-&-&-&-&-&-&-&-&-&-&-\\
96-307&00:07&      46.43& 1.002& 0.00&1324&178&-&-&-&-&-&-&-&-&-&-&-&-&-&-&-\\
96-308&08:12&      36.63& 1.336& 0.01&2520&405& 1.00&1&-&-&-&-&-&-&-&-&-&-&-&-&-\\
96-309&09:06&      27.92& 1.037& 0.04&1154&199& 0.00&1&-&-&-&-&-&-&-&-&-&-&-&-&-\\
 &&&& &&&& &&&& &&&& &&&& &\\
\hline
\end{tabular}
\end{sidewaystable}

\clearpage

\begin{sidewaystable}
\tabcolsep1.5mm
\tiny
\hspace{-2cm} 
\begin{tabular}{|r|r|r|r|ccc|ccc|ccc|ccc|ccc|ccc|}
\hline
&&&& &&&& &&&& &&&& &&&& &\\
\mc{1}{|c|}{Date}&
\mc{1}{|c|}{Time}&
\mc{1}{|c|}{$D_{Jup}$}&
\mc{1}{|c|}{$\Delta $t}&
{\scriptsize $ f_{noi,AC21}$}&{\scriptsize AC}&{\scriptsize AC}&{\scriptsize $ f_{noi,AC22}$}&
{\scriptsize AC}&{\scriptsize AC}&{\scriptsize $ f_{noi,AC23}$}&{\scriptsize AC}&
{\scriptsize AC}&{\scriptsize $ f_{noi,AC24}$}&{\scriptsize AC}&{\scriptsize AC}&
{\scriptsize $ f_{noi,AC25}$}&{\scriptsize AC}&{\scriptsize AC}&{\scriptsize $ f_{noi,AC26}$}&
{\scriptsize AC}&{\scriptsize AC}\\
&&[$\rm R_J$]&\mc{1}{c|}{[d]}&
&$21^{\ast}$&$31^{\ast}$&
&22&32&
&23&33&
&24&34&
&25&35&
&26&36\\
&&&& &&&& &&&& &&&& &&&& &\\
\hline
&&&& &&&& &&&& &&&& &&&& &\\
96-310&04:59&      20.16& 0.828& 0.07&1480&98&-&-&2&-&-&-&-&-&-&-&-&-&-&-&-\\
96-311&00:12&      12.25& 0.801& 0.25&2838&166& 0.00&1&1& 0.00&1&-&-&-&-&-&-&-&-&-&-\\
96-312&01:13&      11.65& 1.042& 0.74&466&41& 0.71&$\,$7$^{\ast}$ &2& 0.00&$\,$2$^{\ast}$ &2& 0.00&1&-&-&-&1&-&-&-\\
96-314&00:45&      30.37& 1.980& 0.92&119&5& 0.54&$\,$13$^{\ast}$ &-& 1.00&$\,$1$^{\ast}$ &2& 0.00&1&-& 0.00&1&1&-&-&-\\
96-327&17:44&      85.63& 13.70& 0.38&358&54& 0.00&1&-&-&-&-&-&-&-&-&-&-&-&-&-\\
 &&&& &&&& &&&& &&&& &&&& &\\
96-330&06:08&      88.24& 2.516& 0.75&14&1&-&-&-&-&-&-&-&-&1&-&-&-&-&-&-\\
96-339&16:35&      82.84& 9.435& 0.00&861&68&-&-&-&-&-&-&-&-&-&-&-&-&-&-&-\\
96-339& 17:00 & 82.80 & 0.017 &\mc{3}{|c|}{---------------------}&\mc{3}{|c|}{---------------------}&\mc{3}{|c|}{---------------------}
&\mc{3}{|c|}{---------------------}&\mc{3}{|c|}{---------------------}&\mc{3}{|c|}{---------------------}\\
96-349&00:17&      49.16& 9.321& 0.00&384&127&-&-&-&-&-&-&-&-&-&-&-&-&-&-&-\\
96-350&20:31&      36.91& 1.843& 0.00&50323$^\$$&9994$^\$$& 0.00&$\,$1$^{\ast}$ &-&-&-&-&-&-&-&-&-&-&-&-&-\\
 &&&& &&&& &&&& &&&& &&&& &\\
96-350&22:38&      36.25& 0.088& 0.00&17091$^\$$&1791$^\$$& 0.00&$\,$1$^{\ast}$ &-&-&-&-&-&-&-&-&-&-&-&-&-\\
96-351&00:24&      35.69& 0.073& 0.00&455$^\$$&85$^\$$&-&-&-&-&-&-&-&-&-&-&-&-&-&-&-\\
96-352&23:57&      18.08& 1.980& 0.05&81422$^\$$&10383$^\$$&-&-&-&-&-&-&-&-&-&-&-&-&-&-&-\\
96-353&06:33&      15.29& 0.275& 0.23&1133$^\$$&65$^\$$&-&-&1&-&-&-&-&-&-&-&-&-&-&-&-\\
96-354&06:27&      9.370& 0.995& 0.48&5957$^\$$&436$^\$$& 0.50&2&1& 0.00&2&-&-&-&-&-&-&-&-&-&-\\
 &&&& &&&& &&&& &&&& &&&& &\\
96-355&00:47&      15.55& 0.763& 0.96&104$^\$$&3$^\$$& 0.33&$\,$15$^{\ast}$ &2& 0.00&1&1& 0.00&1&-& 0.00&1&-&-&-&-\\
96-360&19:31&      54.22& 5.780& 0.74&67$^\$$&2$^\$$& 1.00&4&-&-&-&-&-&-&-&-&-&-&-&-&-\\
96-365&04:07&      67.34& 4.358& 0.88&11$^\$$&-&-&-&-&-&-&-&-&-&-&-&-&-&-&-&-\\
\hline 
\mc{4}{|l|}{}& &&&& &&&& &&&& &&&& &\\
\mc{4}{|l|}{Events (counted)}&- &228730$^{\dagger}$ &37923$^{\dagger}$ &- &113&16&- &9&11&- &4&7&- &2&3&- &0&-\\[1.5ex]
\mc{4}{|l|}{Impacts (complete data)}&-&1807&3451&-&39&16&-&8&11&-&6&7&-&3&3&-&2&-\\[1.5ex]
\mc{4}{|l|}{All events(complete data)}& 0.28&2512&3451& 0.53&83&16& 0.20&10&11& 0.00&6&7& 0.00&3&3& 0.00&2&-\\[1.5ex]
\hline
\end{tabular}
\\
\parbox{19cm}{
{\footnotesize $\ast$: AR2 to AR6: The complete data set was transmitted for only a fraction of all 
particles detected in this amplitude range and time interval. $ f_{noi}$ has been estimated from 
the data sets transmitted. AR1: data transmission is always incomplete in this amplitude range.} \\
{\footnotesize $\dagger$: Due to an unknown number of accumulator overflows before day 96-339 these
 numbers are lower limits for the true numbers of events.} \\
{\footnotesize \$: After day 96-348 accumulator overflows are fully taken into account and the
numbers given are the true numbers of detected events.}
}
\end{sidewaystable}

\setcounter{table}{2}
\begin{table}[ht]
\caption{\label{sec_criteria} Criteria for the identification of secondary 
ejecta grains in the data set.
}
%\small
  \begin{tabular}{cccccc}
   \hline
   \hline \\[-2.0ex]
%Titelzeile
Satellite & 
Orbit  & 
Rotation & 
Impact & 
Class~2  & 
Altitude   
%$\rm R_{sat}$ & 
%Hill  
\\
         & 
  &  
angle range   & 
velocity     & 
denoised   & 
limit [km]    
%   [km]  & 
%radius  &   
              \\[0.7ex]
\hline \\[-2.0ex]
Ganymede & 
G1,G2  & 
$\rm 180^{\circ} \leq ROT \leq 360^{\circ}$ & 
not restricted               &  
no        & 
39,525 
%$\rm 15\,R_{sat}$ &
%2,635   & 
%$\rm 12.0\,R_{sat}$  
\\
Callisto & 
C3     & 
$\rm 180^{\circ} \leq ROT \leq 360^{\circ}$ & 
$v \rm  \leq 10\,km\,s^{-1}$ &  
no        & 
48,180 
%$\rm 20\,R_{sat}$ &
%2,409   & 
%$\rm 20.7\,R_{sat}$  
\\
Europa   & 
E4 & 
$\rm 180^{\circ} \leq ROT \leq 360^{\circ}$ & 
not restricted               & 
 yes       & 
23,400 
%$\rm 15\,R_{sat}$ &
%1,560   & 
%$\rm  8.8\,R_{sat}$   
\\[0.7ex] 
\hline
\end{tabular}\\[1.5ex]
\end{table}

\begin{table*}[hb]
\caption{\label{denoise}
Criteria to identify noise events in class~2 during the E4 fly-by. 
Noise events in the lowest amplitude range (AR1) fulfill at least one 
of the criteria listed below 
(see Paper~I for a definition of the parameters), whereas 
noise events in the higher amplitude ranges fulfil all criteria 
listed for AR2 to AR6.
}
{\small
  \begin{tabular*}{15cm}{lcc}
   \hline
   \hline \\[-2.5ex]
Charge Parameter  &  AR1                         & AR2 to AR6   \\
\hline  
Entrance grid amplitude                  &  PA $\geq$ 9                    &    ---       \\
Channeltron amplitude                    &  ---                            & CA $\leq$ 2  \\
Target amplitude minus iongrid amplitude &  (EA$-$IA) $\leq$ 0 or          & (EA$-$IA) $\leq$ 0 or \\ 
                                         &  (EA$-$IA) $\geq$ 7             & (EA$-$IA) $\geq$ 7 \\
EA risetime                                      & ET $\leq$ 9 or ET = 15  &    ---       \\
IA risetime                                      & IT $\leq$ 8                     &    ---       \\
\hline
\hline
\end{tabular*}\\[1.5ex]
}
\end{table*}

 \tabcolsep0.07cm
 \renewcommand{\arraystretch}{0.1}

 \begin{sidewaystable}
\tabcolsep1.2mm
\tiny
\vbox{
\hspace{-2cm}
\begin{minipage}[t]{22cm}
 {\bf Table 5:}
 DPF data: No., impact time,  TEV 
 CLN, AR, SEC, IA, EA, CA, IT, ET, 
 EIT, EIC, ICC, PA, PET, EVD, ICP, ECP, CCP, PCP, HV and 
 evaluated data: LON, LAT, $\rm D_{Jup}$ (in $\rm R_J$; Jupiter radius $\rm R_J = 71492\,km$), 
rotation angle (ROT), instr. pointing ($\rm S_{LON}$, $\rm S_{LAT}$), speed ($v$, 
in $\rm km\,sec^{-1}$), 
 speed error factor (VEF), mass ($m$, in grams) and mass error factor (MEF).
\end{minipage}
}

\bigskip

\hspace{-2cm}  
\begin{tabular*}{23cm}  {%@{\extracolsep{\fill}}                                 
rrrrrrrrrrrrrrrcrrrrcrrrrrrrrrrrrrrrrrrrrr}                                     
 \mc{1}{c}{No.}&               \mc{1}{c}{IMP. DATE}&         \mc{1}{c}{ TEV }&             \mc{1}{c}{C}&                 
 \mc{1}{c}{AR}&                \mc{1}{c}{S}&                 \mc{1}{c}{IA}&                \mc{1}{c}{EA}&                
 \mc{1}{c}{CA}&                \mc{1}{c}{IT}&                \mc{1}{c}{ET}&                \mc{1}{c}{E}&                 
 \mc{1}{c}{E}&                 \mc{1}{c}{I}&                 \mc{1}{c}{PA}&                \mc{1}{c}{P}&                 
 \mc{1}{c}{E}&                 \mc{1}{c}{I}&                 \mc{1}{c}{E}&                 \mc{1}{c}{C}&                 
 \mc{1}{c}{P}&                 \mc{1}{c}{HV}&                \mc{1}{c}{LON}&               \mc{1}{c}{LAT}&               
 \mc{1}{c}{$\rm D_{Jup}$}&     \mc{1}{c}{ROT}&               \mc{1}{c}{$\rm S_{LON}$}&     \mc{1}{c}{$\rm S_{LAT}$}&     
 \mc{1}{c}{V}&                 \mc{1}{c}{VEF}&               \mc{1}{c}{M}&                 \mc{1}{c}{MEF}\\              
 &                             &                             &                             \mc{1}{c}{L}&                 
 &                             \mc{1}{c}{E}&                 &                             &                             
 &                             &                             &                             \mc{1}{c}{I}&                 
 \mc{1}{c}{I}&                 \mc{1}{c}{C}&                 &                             \mc{1}{c}{E}&                 
 \mc{1}{c}{V}&                 \mc{1}{c}{C}&                 \mc{1}{c}{C}&                 \mc{1}{c}{C}&                 
 \mc{1}{c}{C}&                 &                             &                             &                             
 &                             &                             &                             &                             
 &                             &                             &                             \\                            
 &                             &                             &                             \mc{1}{c}{N}&                 
 &                             \mc{1}{c}{C}&                 &                             &                             
 &                             &                             &                             \mc{1}{c}{T}&                 
 \mc{1}{c}{C}&                 \mc{1}{c}{C}&                 &                             \mc{1}{c}{T}&                 
 \mc{1}{c}{D}&                 \mc{1}{c}{P}&                 \mc{1}{c}{P}&                 \mc{1}{c}{P}&                 
 \mc{1}{c}{P}&                 &                             &                             &                             
 &                             &                             &                             &                             
 &                             &                             &                             \\                            
&&&&& &&&&& &&&&& &&&&& &&&&& &&&&& &\\                                         
\hline                                                                          
&&&&& &&&&& &&&&& &&&&& &&&&& &&&&& &\\                                         
  2884&96-098 17:36:21&259&3&3& 54&20&24& 5& 6& 4& 6&0&1&44& 0&1&0&1&0&1&2&
276.0&  0.0&262.424& 14&271& 52&36.5& 1.6&$ 2.0\cdot 10^{-13}$&   6.0\\
  2887&96-148 11:14:13& 22&3&2& 34&13&21& 2& 9& 4& 6&0&1&41& 0&1&0&1&0&1&2&
279.8& -0.1&175.563& 42&296& 37&12.7& 1.9&$ 2.2\cdot 10^{-12}$&  10.5\\
  2889&96-148 23:52:33& 22&2&2&  5& 8&12& 0& 9&10& 8&0&0&36& 0&1&0&1&0&1&2&
279.8& -0.1&173.949& 83&307&  6&18.3& 1.6&$ 1.1\cdot 10^{-13}$&   6.0\\
  2897&96-165 23:03:47&259&2&2&166&11&11& 4&10&11& 0&1&1& 3&31&1&0&1&0&1&2&
281.0& -0.1&107.912&217&212&-40& 9.7& 1.9&$ 8.5\cdot 10^{-13}$&  10.5\\
  3367&96-178 14:13:39& 22&2&2&112&13&19&18&15&15& 0&1&1&38&29&1&0&1&0&1&2&
281.5&  0.0& 21.388&293&229& 17& 2.5& 1.6&$ 2.1\cdot 10^{-10}$&   6.0\\
&&&&& &&&&& &&&&& &&&&& &&&&& &&&&& &\\                      \hline                                                      
  3419&96-179 04:23:00& 22&2&2& 94& 8&13& 1&11&13&12&0&0&19&24&1&0&1&0&1&2&
281.5&  0.0& 15.774&318&239& 36& 7.4& 1.6&$ 1.5\cdot 10^{-12}$&   6.0\\
  3447&96-179 06:28:55&  6&3&2&128& 8&12&14&13&14& 8&0&1&39&31&1&0&1&0&1&2&
281.5&  0.0& 15.016&270&227&  0& 2.7& 1.6&$ 3.4\cdot 10^{-11}$&   6.0\\
  3450&96-179 06:31:39&  2&3&2&150& 8&11& 4&12&13& 6&0&1& 4&30&1&0&1&0&1&2&
281.5&  0.0& 14.999&239&230&-25& 5.9& 1.6&$ 2.3\cdot 10^{-12}$&   6.0\\
  3497&96-180 10:14:41&  8&2&3&168&20&21&15& 7& 9& 7&0&1&37& 0&1&0&1&0&1&2&
281.5&  0.0& 12.378&214&242&-44&12.7& 1.9&$ 4.7\cdot 10^{-12}$&  10.5\\
  3498&96-180 15:47:20&  8&2&4&254&24&23&18&13&14& 0&1&1&47&16&1&0&1&0&1&2&
281.6&  0.0& 14.042& 93&337& -2& 2.0& 1.9&$ 4.9\cdot 10^{-09}$&  10.5\\
&&&&& &&&&& &&&&& &&&&& &&&&& &&&&& &\\                      \hline                                                      
  3499&96-180 20:58:46&  8&2&3&113&19&25& 5& 8&15& 8&0&1&23& 0&1&0&1&0&1&2&
281.6&  0.0& 15.893&291&229& 16& 5.6& 1.6&$ 8.6\cdot 10^{-11}$&   6.0\\
  3500&96-181 00:16:56& 22&2&4&178&28&30&28&11&12& 8&0&1&20& 0&1&0&1&0&1&2&
281.6&  0.0& 17.150&200&255&-51& 2.7& 1.6&$ 7.3\cdot 10^{-09}$&   6.0\\
  3501&96-183 22:29:07&259&3&4& 18&25&29&22&11&10& 6&0&1&47& 0&1&0&1&0&1&2&
281.9& -0.1& 42.430& 65&333& 19& 5.2& 1.9&$ 4.9\cdot 10^{-10}$&  10.3\\
  3502&96-184 02:47:58&259&2&2& 26& 8&12& 0& 8& 0&12&0&0& 0& 0&1&0&1&0&1&2&
281.9& -0.1& 43.758& 53&330& 28&19.0& 1.9&$ 9.1\cdot 10^{-14}$&  10.5\\
  3575&96-230 11:32:35& 22&3&3& 74&22&27&28& 4& 4& 5&0&1&47& 0&1&0&1&0&1&2&
286.3& -0.2&113.162&346&263& 51&52.6& 1.6&$ 9.8\cdot 10^{-14}$&   6.0\\
&&&&& &&&&& &&&&& &&&&& &&&&& &&&&& &\\                      \hline                                                      
  3581&96-231 00:38:13& 22&3&3&139&20&23& 1& 6& 4& 7&0&1&44& 0&1&0&1&0&1&2&
286.3& -0.2&111.990&255&227&-13&36.5& 1.6&$ 1.7\cdot 10^{-13}$&   6.0\\
  3780&96-241 23:55:26&259&3&2&180& 8& 4&17& 8&14&11&0&1& 8&31&1&0&1&0&1&2&
287.1& -0.2& 75.744&197&254&-52&19.0& 1.9&$ 2.5\cdot 10^{-14}$&  10.5\\
  3782&96-242 17:05:45&259&2&2&145&15& 5& 6& 0& 0& 9&0&0& 0& 0&1&0&1&0&1&2&
287.2& -0.2& 72.299&246&225&-19&11.8&11.8&$ 3.6\cdot 10^{-13}$&5858.3\\
  3783&96-242 17:10:49&259&2&2& 70& 8& 9& 9& 8&10& 3&1&1& 4&31&1&0&1&0&1&2&
287.2& -0.2& 72.281&352&266& 52&19.0& 1.9&$ 5.6\cdot 10^{-14}$&  10.5\\
  3784&96-243 01:48:30&259&2&2& 74& 8& 9& 9& 8&10& 3&1&1& 6&31&1&0&1&0&1&2&
287.2& -0.2& 70.481&346&259& 51&19.0& 1.9&$ 5.6\cdot 10^{-14}$&  10.5\\
&&&&& &&&&& &&&&& &&&&& &&&&& &&&&& &\\                      \hline                                                      
  5194&96-250 06:53:06&  8&3&2&107& 8&12& 8& 9& 9& 5&0&1&36& 0&1&0&1&0&1&2&
287.5& -0.2& 19.729&300&227& 23&19.9& 1.6&$ 7.4\cdot 10^{-14}$&   6.0\\
  5261&96-250 13:29:28&  8&3&4&125&25&29&26& 8& 8& 9&0&1&47& 0&5&0&1&1&1&2&
287.5& -0.2& 17.109&274&223&  2&16.0& 1.6&$ 2.1\cdot 10^{-11}$&   6.0\\
  5341&96-250 18:51:01&  2&2&2&189&10&13&14&14&14& 0&1&1&38&15&5&0&1&1&1&2&
287.5& -0.2& 15.059&184&272&-56& 2.5& 1.6&$ 7.4\cdot 10^{-11}$&   6.0\\
  5347&96-250 18:57:16&  2&2&2&106&15& 4& 2& 0& 5&12&0&0& 4& 0&5&0&1&1&1&2&
287.5& -0.2& 15.019&301&228& 24&11.8&11.8&$ 3.1\cdot 10^{-13}$&5858.3\\
  5369&96-250 19:28:24&  8&3&3&133&20&22&17& 6&10& 9&0&1&40& 0&5&0&1&1&1&2&
287.5& -0.2& 14.822&263&223& -6&15.0& 1.6&$ 3.7\cdot 10^{-12}$&   6.0\\
&&&&& &&&&& &&&&& &&&&& &&&&& &&&&& &\\                      \hline                                                      
  5381&96-250 20:17:56&  8&3&5&105&49&49&28&10&15& 8&0&1&47& 0&5&0&1&1&1&2&
287.5& -0.2& 14.518&302&228& 25&11.8&11.8&$ 2.7\cdot 10^{-10}$&5858.3\\
  5388&96-250 20:53:19&  8&3&3& 99&21&25&12& 8& 9& 8&0&1&46& 0&5&0&1&1&1&2&
287.5& -0.2& 14.305&311&231& 31&14.0& 1.6&$ 8.4\cdot 10^{-12}$&   6.0\\
  5392&96-250 21:28:43&  8&3&2&119& 8&10& 6&12&13& 5&0&1& 2&31&5&0&1&1&1&2&
287.5& -0.2& 14.095&283&224&  9& 5.9& 1.6&$ 2.0\cdot 10^{-12}$&   6.0\\
  5412&96-250 23:21:57&  8&2&4& 99&27&29&26&10& 9& 2&1&1&45& 3&5&0&1&1&1&2&
287.5& -0.2& 13.445&311&235& 14& 9.5& 1.7&$ 1.1\cdot 10^{-10}$&   7.6\\
  5452&96-251 02:47:14&  8&2&2&210& 8&12& 7& 9& 9& 6&0&0&36& 0&5&0&1&1&1&2&
287.5& -0.2& 12.376&155&310&-48&19.9& 1.6&$ 7.4\cdot 10^{-14}$&   6.0\\
&&&&& &&&&& &&&&& &&&&& &&&&& &&&&& &\\                      \hline                                                      
  5512&96-251 10:06:03&  8&3&3&118&22&26&14& 5& 5& 6&0&1&46& 0&5&0&1&1&1&2&
287.5& -0.2& 10.854&284&224& 10&40.9& 1.6&$ 2.4\cdot 10^{-13}$&   6.0\\
  5534&96-251 14:35:00&  8&2&2&174&14&20& 0&15&14&12&0&0&22& 0&5&0&1&1&1&2&
287.5& -0.2& 10.667&205&245&-48& 2.5& 1.6&$ 2.9\cdot 10^{-10}$&   6.0\\
  5535&96-251 16:16:06&100&2&2&109&14&20&15&11& 8& 7&0&1&36& 0&5&0&1&1&1&2&
287.5& -0.2& 10.765&297&226& 20&13.8& 1.6&$ 1.7\cdot 10^{-12}$&   6.0\\
  5536&96-251 16:21:09&  8&2&2& 38&14&10& 5&13&15& 0&1&1&12&29&5&0&1&1&1&2&
287.5& -0.2& 10.772& 37&317& 40& 2.3& 1.9&$ 1.3\cdot 10^{-10}$&  10.5\\
  5537&96-251 17:24:51&  8&2&2&134& 9&22& 4& 9& 2&10&0&0&10& 0&5&0&1&1&1&2&
287.5& -0.2& 10.883&262&223& -7&12.7& 1.9&$ 1.3\cdot 10^{-12}$&  10.5\\
&&&&& &&&&& &&&&& &&&&& &&&&& &&&&& &\\                      \hline                                                      
  5538&96-251 17:39:01&  8&3&4&127&24&27&26& 7& 9& 9&0&1&46& 0&5&0&1&1&1&2&
287.5& -0.2& 10.912&271&223&  0&16.0& 1.6&$ 1.3\cdot 10^{-11}$&   6.0\\
  5539&96-251 19:39:21&  8&2&3&153&19&20&17&10& 7& 0&1&1&40&11&5&0&1&1&1&2&
287.5& -0.2& 11.224&235&228&-28& 4.5& 1.9&$ 7.0\cdot 10^{-11}$&  10.5\\
  5540&96-251 21:11:22&  8&2&2&  2& 9&11& 2&15&11&12&0&0&20& 0&5&0&1&1&1&2&
287.5& -0.2& 11.535& 87&333&  1& 6.4& 3.1&$ 2.1\cdot 10^{-12}$&  58.7\\
  5543&96-252 07:27:08&  8&2&2&252& 9&11& 0&14& 5&13&0&0&47& 2&5&0&1&1&1&2&
287.5& -0.2& 14.690& 96&333& -5&11.8& 9.5&$ 3.9\cdot 10^{-13}$&2690.1\\
  5544&96-252 20:46:54& 22&3&4& 10&26&30&23& 9& 8& 5&0&1&47& 0&1&0&1&0&1&2&
287.6& -0.2& 19.873& 76&332& 10&13.8& 1.6&$ 4.2\cdot 10^{-11}$&   6.0\\
&&&&& &&&&& &&&&& &&&&& &&&&& &&&&& &\\                      \hline                                                      
  5550&96-255 01:49:04& 22&3&4&222&29&49&29&12&15& 5&0&1&19& 0&1&0&1&0&1&2&
287.8& -0.2& 38.942&138&323&-38& 2.0& 1.9&$ 5.1\cdot 10^{-08}$&  10.5\\
  5552&96-258 12:00:04& 22&3&4& 25&29&49&28&13&15& 9&0&1&47& 0&1&0&1&0&1&2&
288.2& -0.2& 60.894& 55&326& 27& 2.0& 1.9&$ 5.1\cdot 10^{-08}$&  10.5\\
  5653&96-289 02:09:10& 22&3&2&229&14&21& 1& 9& 5& 8&0&1&42& 0&1&0&1&0&1&2&
291.2& -0.2&108.151&128&331&-31&12.7& 1.9&$ 2.5\cdot 10^{-12}$&  10.5\\
  6848&96-308 08:12:44&  8&2&2&186& 8&13& 0& 9& 9& 6&0&0&36& 0&1&0&1&0&1&2&
292.5& -0.3& 36.621&188&269&-55&19.9& 1.6&$ 8.7\cdot 10^{-14}$&   6.0\\
  7090&96-309 13:55:36&  2&3&2&105& 9&14& 3&10&11& 9&0&1&37& 0&1&0&1&0&1&2&
292.5& -0.3& 26.089&302&232& 25&14.0& 1.6&$ 4.2\cdot 10^{-13}$&   6.0\\
&&&&& &&&&& &&&&& &&&&& &&&&& &&&&& &\\                      \hline                                                      
  7093&96-309 14:00:21& 22&3&2&134&12&19&17&12&13&11&0&1&36& 0&1&0&1&0&1&2&
292.5& -0.3& 26.059&262&227& -7& 4.5& 1.9&$ 2.5\cdot 10^{-11}$&  10.5\\
  7218&96-310 04:59:15&  8&2&5&153&53&55&27&15&13& 0&1&1&22& 4&1&0&1&0&1&2&
292.5& -0.3& 20.156&235&231&-28& 2.5& 1.6&$ 1.7\cdot 10^{-07}$&   6.0\\
  7285&96-310 11:21:27&  8&2&2& 62&10&14&12&12&14& 0&1&1&38&11&5&0&1&1&1&2&
292.5& -0.3& 17.509&  3&285& 53& 3.8& 1.6&$ 2.1\cdot 10^{-11}$&   6.0\\
  7385&96-310 23:44:37&  8&3&2& 94&10&13& 3&11&12& 8&0&1& 4&31&5&0&1&1&1&2&
292.5& -0.3& 12.421&318&239& 36& 9.2& 1.6&$ 1.2\cdot 10^{-12}$&   6.0\\
  7408&96-311 03:59:24&  8&3&3& 82&23& 1&11&11&15& 6&0&1&45& 0&5&0&1&1&1&2&
292.5& -0.3& 10.919&335&251& 46& 2.3& 1.9&$ 9.3\cdot 10^{-11}$&  10.5\\
\end{tabular*} \end{sidewaystable}                                                     
\clearpage

\begin{sidewaystable}                                                                   
\tabcolsep1.2mm
\tiny 
\hspace{-2cm}
\begin{tabular*}{23cm}  {%@{\extracolsep{\fill}}                                 
rrrrrrrrrrrrrrrcrrrrcrrrrrrrrrrrrrrrrrrrrr}                                     
 \mc{1}{c}{No.}&               \mc{1}{c}{IMP. DATE}&         \mc{1}{c}{ TEV }&             \mc{1}{c}{C}&                 
 \mc{1}{c}{AR}&                \mc{1}{c}{S}&                 \mc{1}{c}{IA}&                \mc{1}{c}{EA}&                
 \mc{1}{c}{CA}&                \mc{1}{c}{IT}&                \mc{1}{c}{ET}&                \mc{1}{c}{E}&                 
 \mc{1}{c}{E}&                 \mc{1}{c}{I}&                 \mc{1}{c}{PA}&                \mc{1}{c}{P}&                 
 \mc{1}{c}{E}&                 \mc{1}{c}{I}&                 \mc{1}{c}{E}&                 \mc{1}{c}{C}&                 
 \mc{1}{c}{P}&                 \mc{1}{c}{HV}&                \mc{1}{c}{LON}&               \mc{1}{c}{LAT}&               
 \mc{1}{c}{$\rm D_{Jup}$}&     \mc{1}{c}{ROT}&               \mc{1}{c}{$\rm S_{LON}$}&     \mc{1}{c}{$\rm S_{LAT}$}&     
 \mc{1}{c}{V}&                 \mc{1}{c}{VEF}&               \mc{1}{c}{M}&                 \mc{1}{c}{MEF}\\              
 &                             &                             &                             \mc{1}{c}{L}&                 
 &                             \mc{1}{c}{E}&                 &                             &                             
 &                             &                             &                             \mc{1}{c}{I}&                 
 \mc{1}{c}{I}&                 \mc{1}{c}{C}&                 &                             \mc{1}{c}{E}&                 
 \mc{1}{c}{V}&                 \mc{1}{c}{C}&                 \mc{1}{c}{C}&                 \mc{1}{c}{C}&                 
 \mc{1}{c}{C}&                 &                             &                             &                             
 &                             &                             &                             &                             
 &                             &                             &                             \\                            
 &                             &                             &                             \mc{1}{c}{N}&                 
 &                             \mc{1}{c}{C}&                 &                             &                             
 &                             &                             &                             \mc{1}{c}{T}&                 
 \mc{1}{c}{C}&                 \mc{1}{c}{C}&                 &                             \mc{1}{c}{T}&                 
 \mc{1}{c}{D}&                 \mc{1}{c}{P}&                 \mc{1}{c}{P}&                 \mc{1}{c}{P}&                 
 \mc{1}{c}{P}&                 &                             &                             &                             
 &                             &                             &                             &                             
 &                             &                             &                             \\                            
&&&&& &&&&& &&&&& &&&&& &&&&& &&&&& &\\                                         
\hline                                                                          
&&&&& &&&&& &&&&& &&&&& &&&&& &&&&& &\\                                         
  7436&96-311 06:06:49&  8&2&3&152&22&22&20&12&11& 0&1&1&47&31&5&0&1&1&1&2&
292.5& -0.3& 10.289&236&231&-27& 2.0& 1.9&$ 2.9\cdot 10^{-09}$&  10.5\\
  7438&96-311 06:49:17&  8&2&2&174& 9&12& 6&12&13& 0&1&1&40&14&5&0&1&1&1&2&
292.5& -0.3& 10.105&205&249&-49& 5.9& 1.6&$ 3.2\cdot 10^{-12}$&   6.0\\
  7439&96-311 09:03:45&  8&3&2&  3&15&20&17&11& 9&10&0&1&37&31&5&0&1&1&1&2&
292.5& -0.3&  9.619& 86&336&  2&12.1& 1.6&$ 2.7\cdot 10^{-12}$&   6.0\\
  7441&96-311 11:53:37&  8&2&2&166& 9&21& 4&15& 9&13&0&0&11&24&5&0&1&1&1&2&
292.5& -0.3&  9.262&217&240&-42& 2.0& 1.9&$ 4.2\cdot 10^{-10}$&  10.5\\
  7442&96-311 16:15:30&  8&3&5&175&49&27&17&15&15& 4&0&1&24& 0&5&0&1&1&1&2&
292.5& -0.3&  9.365&204&250&-49&11.8&11.8&$ 1.1\cdot 10^{-10}$&5858.3\\
&&&&& &&&&& &&&&& &&&&& &&&&& &&&&& &\\                      \hline                                                      
  7444&96-311 16:36:44&  8&3&2&226&10&12& 2&10&10& 8&0&1&37& 0&5&0&1&1&1&2&
292.5& -0.3&  9.409&132&329&-34&16.0& 1.6&$ 2.4\cdot 10^{-13}$&   6.0\\
  7445&96-311 17:46:30&259&3&3& 85&20&21& 4& 8& 7&10&0&1&40& 0&5&0&1&1&1&2&
292.5& -0.3&  9.584&330&248& 44& 9.7& 1.9&$ 9.0\cdot 10^{-12}$&  10.5\\
  7447&96-311 22:44:46& 22&2&4& 45&27&24&14& 5& 4& 0&1&1&23& 2&5&0&1&1&1&2&
292.6& -0.3& 10.825& 27&313& 45&29.8& 1.9&$ 1.2\cdot 10^{-12}$&  10.5\\
  7448&96-312 01:13:24&  8&2&2&254& 9&12& 0&15&11&12&0&0& 3&31&5&0&1&1&1&2&
292.6& -0.3& 11.659& 93&337& -3& 6.4& 3.1&$ 2.5\cdot 10^{-12}$&  58.7\\
  7449&96-312 01:13:24&  8&2&3& 24&20&13&21&10&15& 0&1&1&47&28&5&0&1&1&1&2&
292.6& -0.3& 11.659& 56&331& 26& 4.5& 1.9&$ 3.8\cdot 10^{-11}$&  10.5\\
&&&&& &&&&& &&&&& &&&&& &&&&& &&&&& &\\                      \hline                                                      
  7451&96-312 01:41:43&  8&2&2&228& 8&13& 0& 9& 9& 5&0&0& 1& 0&5&0&1&1&1&2&
292.6& -0.3& 11.829&129&330&-32&19.9& 1.6&$ 8.7\cdot 10^{-14}$&   6.0\\
  7453&96-312 02:31:16&  8&2&2& 89& 8& 8&26& 0& 0& 4&1&1&63& 0&5&0&1&1&1&2&
292.6& -0.3& 12.134&325&243& 40&11.8&11.8&$ 2.0\cdot 10^{-13}$&5858.3\\
  7454&96-312 03:27:54&  8&2&2&204&12&14& 0&13& 9&14&0&0&30& 2&5&0&1&1&1&2&
292.6& -0.3& 12.491&163&305&-53& 8.5& 4.0&$ 2.4\cdot 10^{-12}$& 138.9\\
  7456&96-312 06:03:35&  8&2&2& 38&15& 1& 4& 0&15& 4&0&1&43& 0&5&0&1&1&1&2&
292.6& -0.3& 13.514& 37&336& 38&11.8&11.8&$ 2.0\cdot 10^{-13}$&5858.3\\
  7458&96-312 07:21:27&  8&3&3&210&19&10& 4&10& 1&12&0&1& 5& 1&5&0&1&1&1&2&
292.6& -0.3& 14.042&155&331&-50& 4.5& 1.9&$ 2.1\cdot 10^{-11}$&  10.5\\
&&&&& &&&&& &&&&& &&&&& &&&&& &&&&& &\\                      \hline                                                      
  7460&96-312 08:25:10&  8&2&2&167& 9&22& 3& 8&12& 0&1&1&11&30&5&0&1&1&1&2&
292.6& -0.3& 14.479&215&256&-44&19.0& 1.9&$ 3.7\cdot 10^{-13}$&  10.5\\
  7462&96-312 10:46:43&  8&3&3&235&20&28& 7&14& 2&10&0&1&25& 0&5&0&1&1&1&2&
292.6& -0.3& 15.463&120&350&-25& 2.0& 1.9&$ 5.7\cdot 10^{-09}$&  10.5\\
  7463&96-312 13:43:39&  8&2&5& 90&52&54&30&15&13& 0&1&1&21& 6&5&0&1&1&1&2&
292.6& -0.3& 16.704&323&259& 38& 2.5& 1.6&$ 1.3\cdot 10^{-07}$&   6.0\\
  7464&96-312 19:09:13& 22&3&5& 43&49&52&30&15&14& 5&0&1&47& 0&1&0&1&0&1&2&
292.6& -0.3& 18.983& 30&331& 43& 3.2& 2.0&$ 2.2\cdot 10^{-08}$&  12.5\\
  7465&96-312 23:02:48&  8&2&2&  8&10&13&12&14&15& 0&1&1& 2&31&1&0&1&0&1&2&
292.7& -0.3& 20.593& 79&336&  8& 2.1& 1.6&$ 1.5\cdot 10^{-10}$&   6.0\\
&&&&& &&&&& &&&&& &&&&& &&&&& &&&&& &\\                      \hline                                                      
  7469&96-314 00:45:44& 22&2&4& 62&25&31&11&15& 4& 4&1&1&23& 0&1&0&1&0&1&2&
292.8& -0.3& 30.379&  3&285& 53& 2.0& 1.9&$ 1.9\cdot 10^{-08}$&  10.5\\
  7525&96-330 06:08:15&259&3&4& 48&29&52&13& 7& 6& 5&0&1&47& 0&1&0&1&0&1&2&
294.5& -0.3& 88.246& 23&315& 48&12.7& 1.9&$ 2.3\cdot 10^{-10}$&  10.5\\
  7789&96-350 20:31:24& 22&2&2&148&15& 5& 9& 0& 0& 8&0&0& 0& 0&1&0&1&0&1&2&
296.1& -0.4& 36.909&242&235&-22&11.8&11.8&$ 3.6\cdot 10^{-13}$&5858.3\\
  7804&96-350 22:38:49&  8&2&2&116&10&15& 0& 9&10& 6&0&0&38& 0&1&0&1&0&1&2&
296.1& -0.4& 36.243&287&234& 13&18.3& 1.6&$ 2.3\cdot 10^{-13}$&   6.0\\
  8109&96-352 23:57:20&  8&2&6&169&59&57&27& 6&12& 4&1&1&31& 0&1&0&1&0&1&2&
296.2& -0.4& 18.075&212&249&-44&19.0& 1.9&$ 2.2\cdot 10^{-09}$&  10.5\\
&&&&& &&&&& &&&&& &&&&& &&&&& &&&&& &\\                      \hline                                                      
  8145&96-353 06:33:40& 22&3&2& 71&11&21& 1&14& 4&14&0&1&10& 0&1&0&1&0&1&2&
296.2& -0.4& 15.285&350&274& 53& 2.0& 1.9&$ 5.8\cdot 10^{-10}$&  10.5\\
  8152&96-353 09:23:33& 22&2&3&163&23&27&22&15&14& 0&1&1&10&21&1&0&1&0&1&2&
296.2& -0.4& 14.103&221&244&-38& 2.5& 1.6&$ 3.0\cdot 10^{-09}$&   6.0\\
  8163&96-353 12:55:52& 22&2&3& 91&20&24&19& 7&15& 0&1&1&24&28&5&0&1&1&1&2&
296.2& -0.4& 12.676&322&247& 39& 6.4& 1.6&$ 5.1\cdot 10^{-11}$&   6.0\\
  8184&96-353 19:39:19& 22&3&2& 59&11&14& 7&15&15& 8&0&1&47& 3&5&0&1&1&1&2&
296.2& -0.4& 10.347&  7&297& 53& 2.1& 1.6&$ 2.0\cdot 10^{-10}$&   6.0\\
  8185&96-353 20:43:01& 22&2&6&130&58&13&31& 6& 0& 4&0&1&59& 0&5&0&1&1&1&2&
296.1& -0.4& 10.062&267&232& -2&19.0& 1.9&$ 1.8\cdot 10^{-11}$&  10.5\\
&&&&& &&&&& &&&&& &&&&& &&&&& &&&&& &\\                      \hline                                                      
  8189&96-354 06:28:47&  2&2&2&114&12&21& 5&15&10&11&0&0&10&26&5&0&1&1&1&2&
296.2& -0.4&  9.369&290&234& 15& 2.0& 1.9&$ 6.9\cdot 10^{-10}$&  10.5\\
  8193&96-354 06:51:05&  2&2&2&110&10&12&13&10&10& 2&1&1& 6&31&5&0&1&1&1&2&
296.2& -0.4&  9.422&295&236& 20&16.0& 1.6&$ 2.4\cdot 10^{-13}$&   6.0\\
  8195&96-354 06:52:04&  2&2&3&106&20& 5&14& 8&15&15&0&1&37& 0&5&0&1&1&1&2&
296.2& -0.4&  9.425&301&237& 24& 9.7& 1.9&$ 9.6\cdot 10^{-13}$&  10.5\\
  8197&96-354 06:54:00&  2&3&2&139&10&13&14&11&12& 8&0&1& 2&31&5&0&1&1&1&2&
296.2& -0.4&  9.430&255&233&-12& 9.2& 1.6&$ 1.2\cdot 10^{-12}$&   6.0\\
  8198&96-354 06:54:53&  2&2&2&140& 9&12& 8&12&13& 4&1&1&34&31&5&0&1&1&1&2&
296.2& -0.4&  9.433&253&233&-13& 5.9& 1.6&$ 3.2\cdot 10^{-12}$&   6.0\\
&&&&& &&&&& &&&&& &&&&& &&&&& &&&&& &\\                      \hline                                                      
  8200&96-354 06:55:57&  2&2&2&109&14&20&18&11& 8& 7&0&1&39&31&5&0&1&1&1&2&
296.2& -0.4&  9.436&297&236& 21&13.8& 1.6&$ 1.7\cdot 10^{-12}$&   6.0\\
  8204&96-354 07:00:47&  2&2&2&106&13&19& 9&10&11& 8&0&1& 2&31&5&0&1&1&1&2&
296.2& -0.4&  9.450&301&237& 24& 8.6& 1.6&$ 3.9\cdot 10^{-12}$&   6.0\\
  8205&96-354 07:08:33&  2&2&5&150&49&11&17&15& 0& 5&0&1&58& 0&5&0&1&1&1&2&
296.2& -0.4&  9.473&239&236&-25&11.8&11.8&$ 1.1\cdot 10^{-11}$&5858.3\\
  8208&96-354 12:45:36& 22&2&2& 85&10& 2& 3&12& 2&13&0&0&44&31&5&0&1&1&1&2&
296.2& -0.4& 10.933&330&253& 45& 4.5& 1.9&$ 1.8\cdot 10^{-12}$&  10.5\\
  8209&96-354 13:06:50& 22&3&2& 11&11&25& 4&14&15&13&0&1&63&10&5&0&1&1&1&2&
296.2& -0.4& 11.049& 75&341& 12& 2.0& 1.9&$ 1.1\cdot 10^{-09}$&  10.5\\
&&&&& &&&&& &&&&& &&&&& &&&&& &&&&& &\\                      \hline                 
  8210&96-354 14:53:00& 22&2&2&120&12& 9& 3&13&15& 0&1&1&47&30&5&0&1&1&1&2&
296.2& -0.4& 11.659&281&233&  9& 2.3& 1.9&$ 7.9\cdot 10^{-11}$&  10.5\\
  8211&96-354 14:53:00& 22&3&3&160&20&22&18& 6&10&11&0&1&39& 0&5&0&1&1&1&2&
296.2& -0.4& 11.659&225&242&-35&15.0& 1.6&$ 3.7\cdot 10^{-12}$&   6.0\\
  8212&96-354 15:56:42& 22&2&2&131&11&49&18& 9&11& 0&1&1&28&20&5&0&1&1&1&2&
296.2& -0.4& 12.045&266&232& -3&12.7& 1.9&$ 1.1\cdot 10^{-11}$&  10.5\\
  8213&96-354 18:04:05& 22&2&2&247&13&22& 8&14&11& 0&1&1&49&30&5&0&1&1&1&2&
296.2& -0.4& 12.850&103&342&-10& 2.0& 1.9&$ 9.5\cdot 10^{-10}$&  10.5\\
  8214&96-354 23:50:55&  8&2&4& 13&26& 7&26& 8& 0& 9&0&1&35&19&5&0&1&1&1&2&
296.2& -0.4& 15.168& 72&341& 14& 9.7& 1.9&$ 3.6\cdot 10^{-12}$&  10.5\\
\end{tabular*} \end{sidewaystable}                                                     
\vfill 

\clearpage

\pagestyle{plain}

 \tabcolsep0.25cm
 \renewcommand{\arraystretch}{1.0}

\setcounter{table}{5}

\begin{table}
\caption{\label{stream_table} Times of the onset (class~3), $180^{\circ}$ shift and 
cessation of the Jupiter dust streams. When no entries are given, either 
no RTS data were obtained (onset in G1 and E4) or strong channeltron noise
prevented the detection of dust impacts completely (cessation in G1).
}
  \begin{tabular*}{9cm}{cccc}
   \hline
   \hline \\[-2.0ex]
%Titelzeile
Orbit& Onset         & $180^{\circ}$ & Cessation     \\
     &  class~3      &  shift        &               \\[0.7ex]
\hline \\[-2.0ex]
G1  &  --             & $178.58 \pm 0.05$ & -- \\
G2  & $228.25\pm 3.9$ & $250.20 \pm 0.05$ & $251.6 \pm 0.1$\\
C3  & $291.77\pm 1.7$ & $310.20 \pm 0.10$ & $311.2 \pm 0.1$\\
E4  &     --          & $353.10 \pm 0.10$ & $354.4 \pm 0.7$\\[0.7ex]
\hline
\end{tabular*}\\[1.5ex]
\end{table}

\pagebreak

%\section{Figures}

\end{document}